\documentclass[12pt]{iopart}

\usepackage{epsf}
\usepackage{iopams}
\usepackage{amsfonts}
\usepackage{amssymb}
\usepackage{bm}
\usepackage{bbm}
\usepackage{nicefrac}
\usepackage{tikz}
\usepackage{xcolor}
\usetikzlibrary{arrows,decorations.pathmorphing}
\usepackage{url}

\definecolor{garrosgreen}{rgb}{0.1, 0.4, 0.1}
\definecolor{dartmouthgreen}{rgb}{0.05, 0.5, 0.06}
\definecolor{duelferred}{rgb}{0.7, 0.2, 0.1}
\definecolor{cambridgeblue}{rgb}{0.1, 0.3, 1.0}
\definecolor{oxfordblue}{rgb}{0.05, 0.2, 0.7}

\usepackage[breaklinks=true,colorlinks=true,linkcolor=blue,%
urlcolor=blue,citecolor=blue]{hyperref}
\usepackage{graphicx}
\usepackage{epsfig}
\usepackage{latexsym}
\usepackage{dcolumn}
\usepackage{pifont}
\usepackage{siunitx}
\usepackage{cite}
\newcommand\myapp{\par
  \setcounter{section}{0}
  \setcounter{subsection}{0}
  \setcounter{figure}{0}
  \setcounter{table}{0}
  \renewcommand\thesection{\Alph{section}}
  \renewcommand\thefigure{\Alph{section}\arabic{figure}}
  \renewcommand\thetable{\Alph{section}\arabic{table}}
}

\usepackage{epsfig}
\usepackage{latexsym}

\usepackage{dcolumn}
\usepackage{pifont}

\makeatletter

\newcommand{\Rmnum}[1]{\expandafter\@slowromancap\romannumeral #1@}

\newcommand{\calH}{\mathcal{H}}
\newcommand{\calE}{\mathcal{E}}
\newcommand{\calO}{\mathcal O}

\newcolumntype{.}{D{x}{}{-1}}

\begin{document}

\newcommand{\addrROLLA}{Department of Physics,
Missouri University of Science and Technology,
Rolla, Missouri 65409, USA}

\title[{Casimir-Polder Interaction: Time--Ordered Versus Covariant
Formalism}]{Close Examination of the
Ground--State Casimir--Polder Interaction: Time--Ordered Versus Covariant
Formalism and Radiative Corrections}
   
\author{C M Adhikari and U D Jentschura}
\address{\addrROLLA \\
{\bf email:} adhikaric@mst.edu, ulj@mst.edu}

\begin{abstract}
The purpose of this paper is twofold.
First, we compare, in detail, 
the derivation of the Casimir-Polder interaction using 
time-ordered perturbation theory,
to the matching of the scattering amplitude 
using quantum electrodynamics.  
In the first case, a total of twelve time-ordered diagrams
need to be considered, while in the 
second case, one encounters only two 
Feynman diagrams, namely, the ladder and 
crossed-ladder contributions.
For ground-state interactions,
we match the contribution of six of 
the time-ordered diagrams
against the corresponding Feynman diagrams,
showing the consistency of the two approaches.
Second, we also examine the leading radiative correction to the 
long-range interaction, which is of relative order $\calO(\alpha^3)$.
In doing so, we uncover logarithmic 
terms, in both the interatomic distance as well as 
the fine-structure constant, in higher-order corrections 
to the Casimir--Polder interaction.
\end{abstract}

\vspace{2pc}


%
%
\section{Introduction}
\label{sec1}
As is well known, the 
ground-state Casimir--Polder long-range 
interaction energy between atoms varies as
$1/R^6$ in the short range limit,
where $R$ is interatomic distance.
Due to retardation, it is of the $1/R^7$ type
in the long-range regime~\cite{Casimir-Polder1948}. 
For excited states, it has recently been shown that there
are long-range tails~\cite{UdCmVdPRL2017} 
as a result of retardation. 

The result for the ground state can be obtained 
in two completely different ways,
namely, {\em (i)}
using a covariant approach, with the $S$ matrix formalism,
matching the scattering 
amplitude against the effective Hamiltonian
{\em (ii)} using so-called time-ordered perturbation 
theory, which actually employs time-independent 
field operators in the derivation
and assigns a different diagram to each 
``time ordering'' of the virtual photon emission
and absorption processes.
In the latter case, one encounters twelve diagrams, 
while in the former, only two. 
It would be beyond the scope of the current 
paper to try to review the vast number of investigations
on calculations of interatomic 
long-range (Casimir--Polder)
potentials following the original paper~\cite{Casimir-Polder1948};
let us briefly mention papers 
on multi-electron systems~\cite{MGA1995, MK1996, ZA1997}, 
relativistic corrections as well as other
fundamental questions~\cite{Pachucki2005,Power2001}, 
and particular aspects of 
excited-state interactions~\cite{ET1995, MRA2015, UVCAN2017, UjVd2017}.

A dedicated comparison of the two approaches
has been missing in the literature. 
One advantage of the Feynman formalism is that
it clarifies, uniquely, how to encircle 
the poles of the atomic polarizability matrix element.
Namely, the matching of the $S$ matrix
to the effective Hamiltonian leads to the Feynman
prescription for encircling the poles.
This realization has been instrumental in the treatment
of excited reference states~\cite{UjVd2017,UdCmVdPRL2017}, 
in which case some states of lower energy can become
resonant, and a definite prescription is needed in order to
encircle the poles correctly.
However, in the time-ordered 
formalism, one integrates the virtual photon 
energies $k_1$ and $k_2$ (we set $\hbar = c = \epsilon_0 = 1$)
from zero to infinity (and avoids the Feynman contour).
One possibility to solve the question of how to 
encircle the poles, is to arrange the terms so that the 
characteristic factor 
\begin{eqnarray}
\label{chf}
\frac{1}{k_1 + k_2} - \frac{1}{k_1 - k_2} 
\end{eqnarray}
appears. This factor allows one to symmetrize the 
integrand, so that the $k_2$ integration then proceeds
from $-\infty$ to $+\infty$.
Finally, one carries out the $k_2$ integration
by principal value.
However, even after this somewhat ad hoc prescription
is implemented, the result is only applicable to 
atoms in their ground states.

We anticipate here the result that for the ground state,
the sum of six time-ordered diagrams without crossed photon lines
is equal to one single ``ladder'' Feynman diagram.
Conversely, those time-ordered diagrams
that contain crossed photon lines,
lead to an equal contribution as the
``crossed'' Feynman diagram.
The universality of the end result of the 
derivation for the ground state means that we 
can take it as a safe basis for the 
calculation of relativistic and radiative corrections.
In fact, this program has been implemented
in Ref.~\cite{Pachucki2005}.
It turns out that it is more convenient to use the length gauge for the 
atom-field interaction,
and the so-called temporal (Weyl) gauge for the 
photon propagator. The Weyl gauge has the advantage
over, say, the Coulomb gauge in ensuring that 
the $00$-timelike component of the 
photon propagator can be ignored, while the 
choice of the length gauge leads to a situation 
where we can fortunately 
ignore the seagull term, proportional to $\vec A^2$, 
which would otherwise have to be included 
in the velocity gauge. 

We organize the paper as follows. In Sec.~\ref{sec2}, we calculate the 
long-range interaction energy of the two-atom 
system using time-ordered perturbation theory.
The covariant formalism,
based on the matching of the 
scattering amplitude, is outlined in Sec.~\ref{sec3}. 
In Sec.~\ref{sec4}, we analyze the 
radiative corrections to the Casimir-Polder interactions. 
Conclusions are  drawn in Sec.~\ref{sec5}.
As already stated, we use natural units with 
$\hbar= c= \epsilon_0=1$, and the electron mass is denoted by $m$.

%
%
\section{Time--Ordered Formalism}
\label{sec2}

In order to write 
the unperturbed Hamiltonian for a system of two neutral hydrogen atoms 
$A$ and $B$ (the generalization to multi-electron atoms is straightforward),
one goes into center-of-mass coordinates and defines the 
relative electron coordinates (with respect to the center-of-mass)
to be $\vec{r}_a$ and $\vec{r}_b$,
with the corresponding canonical momenta $\vec p_a$ and $\vec p_b$.
The unperturbed Hamiltonian is
\begin{eqnarray}\label{unpham}
\widehat{H}_{0}=\frac{ \vec{p}_a^{\,2}}{2 m_a} + V(\vec{r}_a) 
+\frac{ \vec{p}_b^{\,2}}{2 m_b} + V(\vec{r}_b) +\widehat{H}_{F}\,,
\end{eqnarray}
Let the center-of-mass of the two atoms 
(roughly equal to the position vectors
of the nuclei) be denoted as $\vec R_A$ and $\vec R_B$.
Then, if the two atoms are far enough apart such that 
$|\vec{r}_a|, |\vec{r}_b| \ll |\vec{R}_A-\vec{R}_B|$,
the potentials $V(\vec{r}_A)$ and   
$V(\vec{r}_{B})$ in Eq.~(\ref{unpham}) can be approximated as 
\begin{eqnarray}\label{potentialab}
V(\vec{r}_A) = - \frac{\alpha}{|\vec{r}_a|},\;\, \qquad \;\,
V(\vec{r}_{B}) = - \frac{\alpha}{|\vec{r}_b|} \,,
\end{eqnarray}
where $\alpha$ is the fine-structure constant.
Substituting $V(\vec{r}_A)$ and $V(\vec{r}_{B})$ in  
Eq.~(\ref{unpham}), the unperturbed  Hamiltonian of 
the system reads
\begin{eqnarray}\label{unphamiltonian}
\widehat{H}_{0}=\frac{ \vec{p}_a^{\,2}}{2m_a}  
- \frac{\alpha}{|\vec{r}_a|}  
+\frac{ \vec{p}_b^{\,2}}{2m_b} 
- \frac{\alpha}{|\vec{r}_b| }+\widehat{H}_{F}\,.
\end{eqnarray}
The first two terms in Eq.~(\ref{unphamiltonian}) 
stand for the Schr\"{o}dinger-Coulomb 
Hamiltonian $\widehat{H}_{A}$, while  the sum of the third and the 
fourth terms are the  Schr\"{o}dinger-Coulomb Hamiltonian 
$\widehat{H}_{B}$. The electromagnetic field Hamiltonian,
$\widehat{H}_{F}$, is given as
\begin{eqnarray}
\widehat{H}_{F}= \sum_{\lambda =1}^2 
\int \mathrm{d}^3k \, k \, a_{\lambda}^{\dagger}(\vec{k}) 
\, a_{\lambda}(\vec{k})\,.
\end{eqnarray}
Here  $a_{\lambda}^{\dagger}$ and $a_{\lambda}$  
are the usual creation and annihilation operators, 
which satisfy the following commutation relation:
\begin{eqnarray}
\left[ a_{\lambda}(\vec{k})\,,
\,a^{\dagger}_{\lambda'}(\vec{k}')\right]=\delta^{(3)}(\vec{k}-\vec{k}')\,
\delta_{\lambda\lambda'}\,.
\end{eqnarray}
Along with the dipole approximation, the interaction 
Hamiltonian in the so-called length gauge 
of quantum electrodynamics (QED) is approximated as
\begin{eqnarray}
\label{HAB}
\widehat{H}_{AB} \approx -e\,  \vec{r}_a \cdot \vec{E}(\vec{R}_A) 
-\mathrm{e}\, \vec{r}_b\cdot \vec{E}(\vec{R}_B),
\end{eqnarray}
where $\vec{E}(\vec{R}_A)$ and $\vec{E}(\vec{R}_B)$ are the 
(Schr\"{o}dinger--picture, time-independent,
see Ref.~\cite{JeKeAOP2004}) electric field operators.
In writing Eq.~(\ref{HAB}), we implicitly assume that the 
wavelength of the exchanged virtual photon is 
much longer than the dimension of the atom,
so that the electric-field operator can be taken at the 
center-of-mass of the atom.
Furthermore, the electromagnetic interaction of the 
proton is taken into account by using the relative 
coordinates $\vec r_a$  and $\vec r_b$ rather than the 
electron coordinates.
Finally, the electromagnetic-field operators are 
given by 
\begin{eqnarray}
\vec{E}(\vec{R}_A) &=
\sum\limits_{\lambda=1}^2{\displaystyle \int \frac{\mathrm{d}^3 k }
{(2\pi)^{3/2}}}\sqrt{\frac{k}{2}} \, 
\widehat{\epsilon}_{\lambda}(\vec{k}) 
\left[\mathrm{i}\, a_{\lambda}(\vec{k}) 
\mathrm{e}^{\mathrm{i}\vec{k}\cdot  \vec{R}_A }
- \mathrm{i} a^{\dagger}_{\lambda}(\vec{k}) 
\mathrm{e}^{-\mathrm{i}\vec{k}. \vec{R}_A } \right]\,,
\end{eqnarray}
and 
\begin{eqnarray}
\vec{E}(\vec{R}_B) =
\sum\limits_{\lambda=1}^2 {\displaystyle \int \frac{\mathrm{d}^3 k }{(2\pi)^{3/2}}}
\sqrt{\frac{k}{2}} \, \widehat{\epsilon}_{\lambda}(\vec{k}) 
\left[\mathrm{i}\, a_{\lambda}(\vec{k}) \mathrm{e}^{\mathrm{i}\vec{k}
\cdot  \vec{R}_B }- \mathrm{i} a^{\dagger}_{\lambda}(\vec{k}) 
\mathrm{e}^{-\mathrm{i}\vec{k}\cdot \vec{R}_B } \right]\,.
\end{eqnarray}
In terms of the creation, annihilation operators of the field, 
the interaction Hamiltonian of the system becomes
\begin{eqnarray}
\label{HABexpl}
\fl\widehat{H}_{AB} &=& - 
e\,\sum\limits_{\lambda=1}^2 
{\displaystyle \int \frac{\mathrm{d}^3 k }{(2\pi)^{3/2}}}
\sqrt{\frac{k}{2}} 
 \left[ \left(\mathrm{i}\, a_{\lambda}(\vec{k})
\widehat{\epsilon}_{\lambda}(\vec{k}) \mathrm{e}^{\mathrm{i}\vec{k}
\cdot \vec{R}_A }- \mathrm{i} 
a^{\dagger}_{\lambda}(\vec{k})\widehat{\epsilon}_{\lambda}(\vec{k}) 
\mathrm{e}^{-\mathrm{i}\vec{k}. \vec{R}_A } \right)
\cdot \vec{r}_a\right. \nonumber\\
\fl&+&\left.\left(\mathrm{i}\, a_{\lambda}(\vec{k})
\widehat{\epsilon}_{\lambda}(\vec{k}) 
\mathrm{e}^{\mathrm{i}\vec{k}\cdot  \vec{R}_B }- 
\mathrm{i} a^{\dagger}_{\lambda}(\vec{k})
\widehat{\epsilon}_{\lambda}(\vec{k}) 
\mathrm{e}^{-\mathrm{i}\vec{k}\cdot  
\vec{R}_B } \right)\cdot \vec{r}_b \right].
\end{eqnarray}
The reference state $|\phi_0\rangle =
|\phi_{1S,A}, \phi_{1S,B}, 0 \rangle$  has 
both atoms $A$ and $B$ in their ground states and  the 
electromagnetic field in the vacuum state $| 0 \rangle$. 
We here calculate the 
perturbation effect of the interaction Hamiltonian. 
The orthonormality condition for the atomic 
parts of the combined atom$+$field state is 
\begin{eqnarray}
\langle n | m\rangle = \delta_{nm},
\end{eqnarray}
where $\delta_{nm}$ is the Kronecker delta and 
$| n\rangle$ and  $| m\rangle$ are any atomic eigenstates
of the atomic part of the unperturbed Hamiltonian $\widehat{H}_0$,
for either atom $A$ or $B$. In the following,
we reserve the notation $| \sigma \rangle$ for a 
virtual state of atom $A$, while a virtual state of atom $B$
is denoted as $| \rho \rangle$.
It is easy to see that all odd-order perturbations
involving the Hamiltonian~(\ref{HAB}) vanish.
The second-order terms are the sum of self-energy 
effects (when both field operators act on the same atom),
as well as one-photon exchange terms which are relevant
only if one has an energetically degenerate,
or quasi-degenerate, state available in either atom,
which can be reached via a dipole transition~\cite{UVCAN2017};
this is typically the case only when excited reference 
states are involved~\cite{ET1995, MRA2015, UVCAN2017, UjVd2017}.
Thus, we look into the fourth order perturbation,  which reads 
\begin{eqnarray}
\label{E4}
\fl \Delta E^{(4)} =\left< \phi_0\left| \widehat{H}_{AB} 
\frac{1}{(E_0-\widehat{H}_{0})^\prime}
\widehat{H}_{AB} \frac{1}{(E_0-\widehat{H}_{0})^\prime}
\widehat{H}_{AB} \frac{1}{(E_0-\widehat{H}_{0})^\prime}  
\widehat{H}_{AB} \right|\phi_0\right>.
\end{eqnarray}
The prime in the operator $\frac{1}{(E_0-\widehat{H}_{0})^\prime}$
indicates that the reference state is excluded from the spectral 
decomposition of the operator.  
The virtual states which need to be used in
the calculation of the fourth-order perturbation~(\ref{E4})
carry one, and two photons in the electromagnetic field 
modes.

\begin{figure}[t!]
\begin{center}
\includegraphics[width=0.8\linewidth]{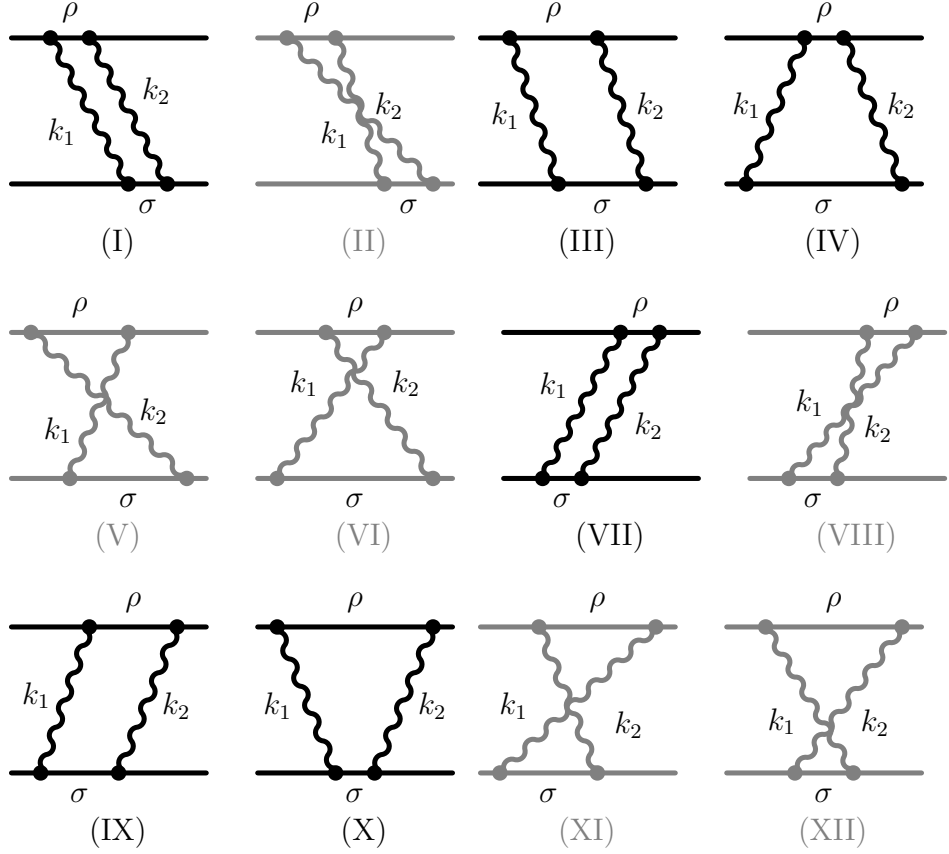}
\caption{\label{figg1}Time-ordered diagrams showing the 
Casimir-Polder interaction 
between two atoms $A$ and $B$. 
The $\rho$ and $\sigma$ lines are the virtual states 
associated with the atom $A$ and the atom $B$. 
The $k_1$ and $k_2$  are  the magnitude of the momenta 
of the photons to the left and to the right of the line respectively.} 
\end{center}
\end{figure}

A Casimir-Polder interaction  between two atoms $A$  and $B$ 
involving two virtual photons results in 
four different types of intermediate states, 
namely, (1) Both atoms are in ground states, 
and two virtual photons are present, 
(2) Only one atom is in the excited state, 
and only one virtual photon is exchanged,
 (3) Both atoms are  in the excited state, but no photon is present, 
and (4) Both atoms are in the excited state, and two photons are 
present~\cite{DPCraig, Akbarsalam}. Thus, the electrons and photons 
can couple in $4\times 3\times 2 \times 1 = 12 $ distinct ways. 
Fig.~\ref{figg1} represents all these 12 
time-ordered sequences of the interaction.

Let us first investigate  the first diagram 
of the Fig.~\ref{figg1}. There are four factors 
which give contributions to the interaction energy, 
namely, emission of $\vec{k}_2$ at $R_B$, 
emission of  $\vec{k}_1$ at $R_B$, absorption of  
$\vec{k}_2$ at $R_A$, and absorption of  
$\vec{k}_1$ at $R_A$.  The corresponding 
fourth-order energy shift reads
\begin{eqnarray}\label{firsttermsimplify}
\fl\Delta  E_I^{(4)}=
e^4 \int \frac{\mathrm{d}^3 k _1}{(2\pi)^3}   
\int \frac{\mathrm{d}^3 k _2}{(2\pi)^3}  \sum_{\lambda_1, \lambda_2} 
\sum_{\rho, \sigma} \frac{k_1k_2}{4}
\, (\mathrm{i}) \, \langle \phi_{1S,A} | 
\widehat{\epsilon}_{\lambda_1}(\vec{k}_1 ) \cdot  \vec{r}_a  
|\rho\rangle \mathrm{e}^{\mathrm{i}\vec{k}_1 
\cdot \vec{R}_A } (- \mathrm{i})\nonumber\\
\fl\qquad\times
 \langle  \phi_{1S,B}| \widehat{\epsilon}_{\lambda_1}(\vec{k}_1 )
 \cdot  \vec{r}_b |\sigma\rangle
  \mathrm{e}^{-\mathrm{i}\vec{k}_1 \cdot  \vec{R}_B }\,(\mathrm{i})
  \langle \rho| 
\widehat{\epsilon}_{\lambda_2} (\vec{k}_2 ) 
\cdot  \vec{r}_a |\phi_{1S,A}  \rangle  \mathrm{e}^{\mathrm{i}\vec{k}_2 
\cdot  \vec{R}_A } (- \mathrm{i} )  \nonumber\\
\fl\qquad\times \langle \sigma| 
\widehat{\epsilon}_{\lambda_2}(\vec{k}_2 ) \cdot \vec{r}_b 
|\phi_{1S,B}  \rangle \mathrm{e}^{-\mathrm{i}\vec{k}_2
 \cdot \vec{R}_B} \,
  \frac{1}{(E_{1S,A}-E_{\rho}-  k_1)
(- k_1-  k_2)
(E_{1S,B}-E_{\sigma}- k_2)} \,,
 \end{eqnarray}
where  $ |\phi_{1S,i}  \rangle$,  $i=A, B\,$ is  a ket associated to 
the ground state  of the atom $i$. 
The summation over the virtual states $| \rho \rangle$
and $| \sigma \rangle$ of atoms $A$ and $B$ 
includes an integral over the continuous spectrum.
Atom $B$ undergoes the transition 
$|1S,B \rangle \to | \sigma \rangle \to | 1S,B \rangle$,
each time under the emission of photons,
while atom $A$ undergoes the transition
$|1S,A \rangle \to | \rho \rangle \to | 1S,A \rangle$,
each time under the absorption of a photon.
We have used Eqs.~(\ref{HAB})---(\ref{HABexpl}).

The polarization vectors 
$ \widehat{\epsilon}_{\lambda_i}(\vec{k}_i)$, with $i = 1,2 $, satisfy 
the following identities,
\begin{eqnarray}
\fl\quad  \widehat{\epsilon}_{\lambda_i}(\vec{k})\cdot 
\widehat{\epsilon}_{\lambda_j}(\vec{k}) = \delta_{\lambda_i\lambda_j}\,,\quad
\vec{k}\cdot  \widehat{\epsilon}_{\lambda_i}&(\vec{k})  = 0\,,
\quad 
\sum\limits_{\lambda_i=1}^2 \widehat{\epsilon}^{\,p}_{\lambda_i}
(\vec{k}_r ) \widehat{\epsilon}^{\,q}_{\lambda_i}(\vec{k}_r ) 
=  \delta^{pq} - \frac{k_{r}^{p}k_{r}^{q}}
{ \vec{k}^{\,2}_r}\,.
\end{eqnarray}
Thus, the contribution to the interaction energy from 
the first diagram reads 
\begin{eqnarray}\label{firsttermfinal}
\fl\Delta E_I^{(4)}&=&e^4
\int \frac{\mathrm{d}^3 k _1}{(2\pi)^3}   
\int \frac{\mathrm{d}^3 k _2}{(2\pi)^3}
\frac{k_1 k_2}{4} \left( \delta^{m r} 
- \frac{k_1^m k_1^r}{k_1^2}\right)
 \left( \delta^{n s } - \frac{k_2^n k_2^s}{k_2^2}\right) 
\mathrm{e}^{\mathrm{i}(\vec{k}_1  +\vec{k}_2 )\cdot
 \vec{R}}  \nonumber\\
\fl&\times&\sum_{\rho, \sigma}
 \frac{ \langle \phi_{1S,A} | x^m |\rho\rangle \langle
 \rho| x^n |\phi_{1S,A}  \rangle \langle \phi_{1S,B} 
| x^r |\sigma\rangle \langle \sigma| x^s |\phi_{1S,B}  \rangle}
{(E_{1S,A}-E_{\rho}- k_1)(-  k_1-  k_2)
(E_{1S,B}-E_{\sigma}-  k_2)}\,,
\end{eqnarray}
where $\vec{R}=\vec{R}_A-\vec{R}_B$ is the internuclear separation.
If we denote a propagator denominator by $ \mathcal{D}_{K}$, 
where $K$ is the roman numeral identifying a diagram in Fig.~\ref{figg1},
then  for diagram (I), we have,
\begin{eqnarray}\label{pd1a}
\mathcal{D}_{{}_{I}}=  (E_{1S,A}-E_{\rho}- k_1)
(- k_1- k_2)(E_{1S,B}-E_{\sigma}-  k_2)\,.
 \end{eqnarray}
Note that in the virtual state in the ``middle'' 
of diagram~${\rm I}$,
both atoms are in the ground state.

The net fourth order energy shift of the system  
is the sum of the contributions 
of all the 12 diagrams. Explicitly,
\begin{eqnarray}\label{12terms}
\fl\Delta E^{(4)}&=&
e^4\int \frac{\mathrm{d}^3 k _1}{(2\pi)^3}   
\int \frac{\mathrm{d}^3 k _2}{(2\pi)^3}   \frac{k_1k_2}{4}
\left( \delta^{m r} - \frac{k_1^m k_1^r}{k_1^2}\right) 
\left( \delta^{n s } - \frac{k_2^n k_2^s}{k_2^2}\right)  
 \mathrm{e}^{\mathrm{i}(\vec{k}_1  +\vec{k}_2 )
\cdot \vec{R}}  
\nonumber\\
\fl &\times& \sum_{\rho, \sigma}  \langle \phi_{1S,A}
 | x^m |\rho\rangle\langle \rho| x^n |\phi_{1S,A}  \rangle
\langle \phi_{1S,B} | x^r |\sigma\rangle \langle \sigma|
x^s |\phi_{1S,B}  \rangle \sum_{j= {\rm I}}^{{\rm XII}} \mathcal{D}_j^{-1}.
\end{eqnarray}

The six diagrams we want to study first are 
${\rm I}$, ${\rm III}$, ${\rm IV}$,
${\rm VII}$, ${\rm IX}$, and ${\rm X}$.
The rationale behind the grouping is that the
photon-lines of the six mentioned time-ordered diagrams do not cross 
(they are the dark-colored in Fig.~\ref{figg1}).
By contrast, photon lines cross in the rest of other six diagrams 
(gray diagrams of Fig.~\ref{figg1}).
Our treatment is inspired by 
Ref.~\cite{DPCraig} but specialized to
the mentioned sets of time-ordered diagrams.

Let us look at the diagram ${\rm III}$,  which is 
the second time-ordered diagram without a photon-line 
crossing (see Fig.~\ref{figg1}).   
The diagram ${\rm III}$ involves the emission of 
a photon with wave vector 
$\vec{k}_2$ at $\vec R_B$, 
the emission of $\vec{k}_1$ at $\vec R_A$,
and the excitation of both atoms. Thus, the propagator denominator 
($\mathcal{D}_{{}_{\rm III}}$) corresponding to the diagram (III)  reads 
\begin{eqnarray}\label{pd3}
\fl \qquad \mathcal{D}_{\rm III}=&  (E_{1S,A}-E_{\rho}-k_1)(
E_{1S,A}-E_{\sigma}+E_{1S,B}-E_{\rho})
(E_{1S,B}-E_{\sigma}- k_2)\,.
\end{eqnarray}
The corresponding energy shift is
\begin{eqnarray}
\label{3rdtermfinal}
\fl\Delta E_{\rm III}^{(4)} &= &
e^4{\displaystyle \int \frac{\mathrm{d}^3 k _1}{(2\pi)^3}   
\int \frac{\mathrm{d}^3 k _2}{(2\pi)^3}  } \frac{k_1k_2}{4}
\Big( \delta^{m r} - \frac{k_1^m k_1^r}{k_1^2}\Big)
 \left( \delta^{n s } - \frac{k_2^n k_2^s}{k_2^2}\right)  
 \mathrm{e}^{\mathrm{i}(\vec{k}_1  +\vec{k}_2 )\cdot 
 \vec{R}} \nonumber\\
\fl&\times&\sum_{\rho, \sigma} 
 \frac{ \langle \phi_{1S,A} | x^m |\rho\rangle
 \langle \rho| x^n |\phi_{1S,A}  \rangle \langle \phi_{1S,B}
 | x^r |\sigma\rangle \langle \sigma| x^s |\phi_{1S,B}  \rangle}
 {(E_{1S,A}-E_{\rho}- k_1)(E_{1S,A}-E_{\sigma}+E_{1S,B}-E_{\rho})
(E_{1S,B}-E_{\sigma}-  k_2)}\,.
\end{eqnarray} 
The propagator denominators $\mathcal{D}_{\rm IV}$,   
$\mathcal{D}_{\rm VII}$, $\mathcal{D}_{\rm IX}$, 
and $\mathcal{D}_{\rm X}$, 
of the diagrams IV, VII, IX, and X  in  
Fig.~\ref{figg1}, respectively, are  given by 
\numparts
\begin{eqnarray}
\fl \qquad  \mathcal{D}_{{}_{IV}}&=& (E_{1S,B}-E_{\sigma}- k_1)
(E_{1S,A}-E_{\rho}+E_{1S,B}-E_{\sigma})(E_{1S,B}-E_{\sigma}- k_2)\,,\\
\fl \qquad   \mathcal{D}_{{}_{VII}}&=&(E_{1S,B}-E_{\sigma}- k_1)
(-k_1- k_2)(E_{1S,A}-E_{\rho}- k_2),\\
\fl  \qquad   \mathcal{D}_{{}_{IX}}&=&(E_{1S,B}-E_{\sigma}- k_1)
(E_{1S,A}-E_{\rho}+E_{1S,B}-E_{\sigma}) (E_{1S,A}-E_{\rho}- k_2), \\
\fl \qquad   \mathcal{D}_{{}_X}&=&(E_{1S,A}-E_{\rho}-k_1)
(E_{1S,A}-E_{\rho}+E_{1S,B}-E_{\sigma})  (E_{1S,A}-E_{\rho}- k_2)\,.
\end{eqnarray}  
\endnumparts
For simplicity, we denote 
$E_{\rho}- E_{1S,A}= E_{A\rho}$ and  
$E_{\sigma}-E_{1S,B}= E_{B\sigma}$. 

Let us now group, simplify,  and then 
assemble the propagator denominators as below:
\numparts
\begin{eqnarray}
\label{EXPd1InIII}
\fl
\frac{1}{\mathcal{D}_{\rm I}} + 
\frac{1}{\mathcal{D}_{\rm III}}
=  \frac{1}{k_1 + k_2} \left(
\frac{-1}{(E_{A\rho}+ E_{B\sigma}) (E_{B\sigma} +k_2)}
+ \frac{-1}{(E_{A\rho}+ E_{B\sigma})(E_{A\rho}+ k_1)} 
\right), \\
\label{EXPd1IV}
\fl\mathcal{D}^{-1}_{\rm IV} 
= \frac{1}{ E_{A\rho}+E_{B\sigma}}
\bigg(\frac{1}{E_{B\sigma}+ k_1}
-\frac{1}{E_{B\sigma}+ k_2}\bigg) \frac{1}{k_1- k_2}\,, \\
\label{EXPd1VIInIX}
\fl
\frac{1}{\mathcal{D}_{\rm VII}} + 
\frac{1}{\mathcal{D}_{\rm IX}}
=  \frac{1}{k_1 + k_2} \left(
\frac{-1}{(E_{A\rho}+ E_{B\sigma})(E_{B\sigma}+ k_1)}
+ \frac{-1}{(E_{A\rho}+ E_{B\sigma})(E_{A\rho}+ k_2)} 
\right), \\
\label{EXPd1X}
\fl\mathcal{D}^{-1}_{\rm X} 
=\frac{1}{E_{A\rho}+E_{B\sigma}} 
\bigg(  \frac{1}{E_{A\rho}+ k_1} 
- \frac{1}{E_{A\rho}+ k_2} \bigg) \frac{1}{k_1-k_2}\,.
\end{eqnarray}
\endnumparts
Adding Eqs.~(\ref{EXPd1InIII})--(\ref{EXPd1X}) and simplifying, 
we obtain for the ``ladder'' (hence the subscript~L) contribution,
\begin{eqnarray}\label{pd1347910}
\fl D^{-1}_{\rm L} &=&
\mathcal{D}^{-1}_{\rm I} +
\mathcal{D}^{-1}_{\rm III} +
D^{-1}_{\rm IV} +
\mathcal{D}^{-1}_{\rm VII} +
\mathcal{D}^{-1}_{\rm IX} +
\mathcal{D}^{-1}_{\rm X} 
\nonumber\\
\fl&=& -\frac{1}{(E_{A\rho}+ E_{B\sigma})} 
\Big( \frac{1}{E_{A\rho}+ k_1}
+\frac{1}{E_{B\sigma}+ k_1}\Big)\Big(\frac{1}{ k_1+ k_2} 
- \frac{1}{ k_1- k_2}\Big)\nonumber\\
\fl \;& & - \frac{1}{(E_{A\rho}+ E_{B\sigma})}
 \Big( \frac{1}{E_{A\rho}+ k_2}
+\frac{1}{E_{B\sigma}+ k_2}\Big) \Big(\frac{1}{ k_1+ k_2} 
+ \frac{1}{ k_1-k_2}\Big)\,.
\end{eqnarray}
We see the characteristic factor~(\ref{chf}) emerge.
Furthermore, we notice that the second term is
equivalent to the first, which implies that the terms
lead to equivalent contributions under the 
photon integral.

The fourth order energy shift due to the six time-ordered 
diagrams I, III, IV, VII, IX, and X, simplifies to
\begin{eqnarray}\label{E4all}
\fl E_{\mathrm{L}}(R)&=&- e^4
 \int \frac{\mathrm{d}^3 k _1}{(2\pi)^3}   \int \frac{\mathrm{d}^3 k _2}{(2\pi)^3}
  \frac{k_1k_2}{4} \left( \delta^{m r} - \frac{k_1^m k_1^r}{k_1^2}\right)
 \left( \delta^{n s } - \frac{k_2^n k_2^s}{k_2^2}\right) 
\mathrm{e}^{\mathrm{i}(\vec{k}_1  +\vec{k}_2 )\cdot \vec{R}}\nonumber\\
\fl&\times& \sum_{\rho, \sigma}\langle \phi_{1S,A} | x^m
 |\rho\rangle \langle \rho| x^n |\phi_{1S,A} \rangle 
  \langle \phi_{1S,B} | x^r |\sigma\rangle
 \langle \sigma| x^s |\phi_{1S,B}\rangle \, \mathcal{D}^{-1}_{L}\nonumber\\
\fl &=& -\frac{ e^4}{18} \int \frac{\mathrm{d}^3 k _1}{(2\pi)^3}  
 \int \frac{\mathrm{d}^3 k _2}{(2\pi)^3} \,
 k_1k_2\,  \delta^{mn}\delta^{rs}  \left( \delta^{m r} - \frac{k_1^m k_1^r}{k_1^2}\right) 
\left( \delta^{n s } - \frac{k_2^n k_2^s}{k_2^2}\right) 
 \mathrm{e}^{\mathrm{i}(\vec{k}_1  +\vec{k}_2 )\cdot \vec{R}} \nonumber\\
\fl &\times&\sum_{\rho, \sigma} \sum_{j,\ell}
 \langle \phi_{1S,A} | x^j |\rho\rangle 
 \langle \rho| x^j |\phi_{1S,A}
   \rangle\langle \phi_{1S,B} | x^{\ell} |\sigma\rangle
 \langle \sigma| x^{\ell} |\phi_{1S,B}  \rangle\nonumber\\
\fl & \times& 
  \frac{(E_{A\rho}+ E_{B\sigma}+2 k_1)}{(E_{A\rho}+ E_{B\sigma})
  (E_{B\sigma}+ k_1)(E_{A\rho}+ k_1)} \left(\frac{1}{ k_1+ k_2} 
- \frac{1}{ k_1- k_2} \right)\,,
  \end{eqnarray}  
where we have used the following identity
\begin{eqnarray}\label{identityII}
\fl  \qquad \sum_{i,j} \langle\phi_{1S,A}|x^i| \rho \rangle
\langle \rho | x^j| \phi_{1S,A} \rangle
= \frac{\delta^{ij}}{3}\sum_s 
\langle \phi_{1S,A} | x^s | \rho \rangle
 \langle \rho | x^s | \phi_{1S,A}\rangle,
\end{eqnarray}
which is valid for any $S$ state.  Using the identity
$\int \mathrm{d}^3k = \int_0^{\infty} k^2 \mathrm{d}k \,
\int \mathrm{d} \Omega$, where
$\mathrm{d} \Omega = \sin\theta \mathrm{d}\theta \mathrm{d} \phi$,
the angular part of Eq.~(\ref{E4all}) can be integrated as 
\begin{eqnarray}\label{angularp}
\fl\int_0^{\pi} \sin\theta \mathrm{d}\theta \int_0^{2\pi} 
\mathrm{d} \phi  \Big( \delta^{m r} - \frac{k_1^m k_1^r}{k_1^2}\Big)
\mathrm{e}^{\mathrm{i}\vec{k}_1 \cdot  \vec{R}} \nonumber\\
\fl \qquad =4\pi \left[ \left( \delta^{m r} - \frac{R^mR^r}{R^2}\right)
\frac{\sin k_1R}{ k_1R}\
+ \left( \delta^{m r} -3 \frac{R^mR^r}{R^2}\right)
\left(\frac{\cos k_1R}{(k_1R)^2} 
-\frac{\sin k_1R}{(k_1R)^3}\right)\right]\,.
\end{eqnarray}
With the help of  Eq.~(\ref{angularp}), 
Eq.~(\ref{E4all}) can be re-expressed as  
\begin{eqnarray}\label{deltaE4net}
\fl E_{\mathrm{L}}(R) 
= \frac{- e^4}{36 \pi^4} \sum_{\rho, \sigma} \sum_{j,\ell}
\frac{ \langle \phi_{1S,A} | x^j |\rho\rangle \langle \rho| x^j |\phi_{1S,A}  \rangle
 \langle \phi_{1S,B} | x^{\ell} |\sigma\rangle \langle \sigma| x^{\ell}
|\phi_{1S,B}  \rangle}{(E_{A\rho}+ E_{B\sigma})}   
\int\limits_0^{\infty} \mathrm{d} k_1 A^{mr}(k_1R)
\nonumber\\
\fl\quad\times
\frac{k_1^3 \,(E_{A\rho}+ E_{B\sigma}+2 k_1)}{ (E_{B\sigma}+ k_1)
(E_{A\rho}+ k_1)}\,   \delta^{mn}\; \delta^{rs} 
\left(\int_0^{\infty} \mathrm{d} k _2\, k_2^3\; \frac{A^{ns}(k_2R)}{(k_1+k_2)} - 
\int_0^{\infty} \mathrm{d} k _2\, k_2^3\, \frac{A^{ns}(k_2R)}{(k_1-k_2)} \right), 
\end{eqnarray} 
where 
\begin{eqnarray}
A^{ns}(x)= \left(\delta^{ns} - \frac{R^nR^s}{R^2}\right)\frac{\sin x}{ x}
+ \left( \delta^{ns} -3 \frac{R^nR^s}{R^2}\right)
\left(\frac{\cos x}{x^2} - \frac{\sin x}{x^3}\right) 
\end{eqnarray}
is an even function of $x$, which allows us to 
extend the integration limit from $k_2=-\infty$ to $k_2=+\infty$. 
Consequently, we have 
\begin{eqnarray}\label{deltaE4k2pm}
\fl E_{\mathrm{L}}(R) 
&=& - \frac{  e^4}{72 \pi^4 } \sum_{\rho, \sigma}\sum_{j,\ell}
\frac{ 1}{E_{A\rho}+ E_{B\sigma}}  
\langle \phi_{1S,A} | x^j |\rho\rangle
\langle \rho| x^j |\phi_{1S,A}  \rangle
\langle \phi_{1S,B} | x^{\ell} |\sigma\rangle
 \langle \sigma| x^{\ell} |\phi_{1S,B} \rangle\nonumber\\
\fl &\times&\int_0^{\infty}  \mathrm{d} k _1\, k_1^3 \delta^{mn}\, \delta^{rs}
\, A^{mr}(k_1R) 
\,\frac{(E_{A\rho}+ E_{B\sigma}+2 k_1)}
 { (E_{B\sigma} +k_1)(E_{A\rho}+k_1)}
\int_{-\infty}^{\infty} \mathrm{d} k _2\, k_2^3\,
 \frac{A^{ns}(k_2R)}{(k_1+k_2)} \,,
\end{eqnarray} 
where we have used the symmetry of the 
integrand in order to extend the integration limits
to the interval $-\infty < k_2 < \infty$.
The $k_2$-integral has a pole of order one at $k_2=-k_1$.
Strictly speaking, the $k_2$ integral in Eq.~(\ref{deltaE4k2pm})
is not uniquely defined, 
and its value depends on the integration prescription.
In the following, we shall implicitly assume that a 
principal-value prescription is indicated.
Let $k_2R=x$ and $k_1R=x_1$.
Then  the $k_2$-integral can be written as 
\begin{eqnarray}\label{Integralk2witheta}
\fl\int_{-\infty}^{\infty} \mathrm{d} k _2\, k_2^3\,
 \frac{A^{ns}(k_2R)}{(k_1+k_2)}\nonumber\\
\fl\qquad=\frac{1}{R^3}  \left(\delta^{ns} - \frac{R^nR^s}{R^2}\right)
 \lim_{\eta\to 0}\left\{\int_{-\infty}^{\infty} \mathrm{d}x
 \frac{x^2}{x+x_1} \frac{\mathrm{e}^{\mathrm{i}x -\eta|x|} }{2\mathrm{i}}
-\int_{-\infty}^{\infty} \mathrm{d}x \frac{x^2}{x+x_1} 
 \frac{\mathrm{e}^{-\mathrm{i}x-\eta|x|}}{2\mathrm{i}}\right\} \nonumber\\
 \fl \qquad+ \frac{1}{R^3}  \left(\delta^{ns} - 3\frac{R^nR^s}{R^2}\right) 
 \lim_{\eta\to 0}\left\{\int_{-\infty}^{\infty} \mathrm{d}x \frac{x}{x+x_1}
\frac{\mathrm{e}^{\mathrm{i}x-\eta|x|}}{2}
+\int_{-\infty}^{\infty} \mathrm{d}x \frac{x}{x+x_1} 
\frac{\mathrm{e}^{-\mathrm{i}x-\eta|x|}}{2}\right\}\nonumber\\
\fl\qquad+\frac{1}{R^3}  \left(\delta^{ns} - 3\frac{R^nR^s}{R^2}\right) 
 \lim_{\eta\to 0}\left\{ \int_{-\infty}^{\infty} \mathrm{d}x \frac{1}{x+x_1} 
\frac{\mathrm{e}^{\mathrm{i}x-\eta|x|}}{2\mathrm{i}}
 -\int_{-\infty}^{\infty} \mathrm{d}x \frac{1}{x+x_1} 
\frac{\mathrm{e}^{-\mathrm{i}x-\eta|x|}}{2\mathrm{i}}\right\} \,,
\end{eqnarray}
where we have introduced a convergence factor 
$\mathrm{e}^{-\eta|x|}$ to make our integrands divergence-free.  

\begin{figure}[t!]
\begin{center}
\includegraphics[width=0.8\linewidth]{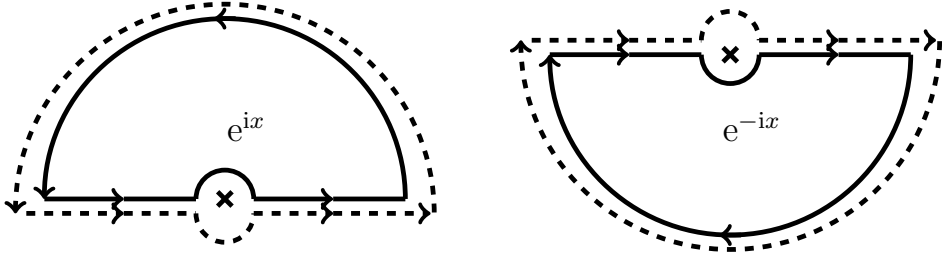}
\caption{ \label{figg2}
Complex integration contours to calculate the principal value of 
integrals in Eq.~(\ref{Integralk2witheta}). }
\end{center}
\end{figure}

It is quite surprising that the principal-value integrals
in Eq.~(\ref{Integralk2witheta}) can be 
evaluated using Cauchy's residue theorem (see Fig.~\ref{figg2}).
One identifies the principal-value evaluation
with a symmetric encircling of the pole,
on a half-circle either above or below the 
pole in the complex plane, and then closes the 
contour in the appropriate half of the complex plane,
as dictated by the functional form of the 
exponential [$\exp(\mathrm{i}x ) \to$ upper half,
$\exp(-\mathrm{i}x ) \to$ lower half].
We finally take the limit $\eta\rightarrow 0$ at the end, which yields
\begin{eqnarray}\label{Integralk2withetaB000}
\fl\qquad\int_{-\infty}^{\infty} 
\mathrm{d} k _2\, k_2^3\ \frac{A^{ns}(k_2R)}{(k_1+k_2)}
&& = \frac{1}{R^3}  \left(\delta^{ns} - \frac{R^nR^s}{R^2}\right) \pi x_1^2 \, \cos x_1 
- \frac{1}{R^3}  \left(\delta^{ns} - 3\frac{R^nR^s}{R^2}\right)\nonumber\\
&& \times\left( \pi x_1 \, \sin x_1
- \pi\, \cos x_1\right) \,.
\end{eqnarray}
Rearranging   Eq.~(\ref{Integralk2withetaB000}) 
and replacing the assumed variable $x_1$ by its value 
$x_1= k_1R$, one obtains
\begin{eqnarray}\label{k2integral}
\fl\qquad\int_{-\infty}^{\infty} \mathrm{d} k _2\, k_2^3 \frac{A(k_2R)}{(k_1+k_2)} 
&&= \pi k_1^3  \Bigg[ \left( \delta^{ns} - \frac{R^nR^s}{R^2}\right)
\frac{\cos k_1R}{ k_1R} - \left(\delta^{ns} -3 \frac{R^nR^s}{R^2}\right)\nonumber\\
&&\times\left(\frac{\sin k_1R}{(k_1R)^2} +\frac{\cos k_1R}{(k_1R)^3}\right)\Bigg]\,,
\end{eqnarray}
which we  substitute to
Eq.~(\ref{deltaE4k2pm}) and  carry out the algebra to get 
\begin{eqnarray}\label{deltaE4net2}
\fl E_{\mathrm{L}}(R)
&&=-\,\frac{ e^4}{72 \pi^3} \sum_{\rho, \sigma} \sum_{j,\ell}  
\frac{ \langle \phi_{1S,A} | x^j |\rho\rangle
 \langle \rho| x^j |\phi_{1S,A}  \rangle   
 \langle \phi_{1S,B} | x^{\ell} |\sigma\rangle
\langle \sigma| x^{\ell} |\phi_{1S,B}  \rangle}
{(E_{A\rho}+ E_{B\sigma})} 
\nonumber\\
\fl&&\times\int_0^{\infty} \mathrm{d} k _1 \,  k_1^6  \,
\frac{ (E_{A\rho}+ E_{B\sigma}+ 2 k_1) }{ (E_{B\sigma}
+ k_1)(E_{A\rho}+k_1)} \left[ \frac{\sin 2k_1R}{(k_1R)^2} 
- \frac{2\, \sin ^2k_1R}{(k_1R)^3} 
+ \frac{2\, \cos ^2k_1R}{(k_1R)^3}\right.\nonumber\\
\fl&& \left. - \frac{5\, \sin 2k_1R}{(k_1R)^4 } 
-   \frac{6\, \cos ^2k_1R}{(k_1R)^5 } + \frac{6\, \sin ^2k_1R}{(k_1R)^5 } 
+  \frac{3\, \sin 2k_1R}{(k_1R)^6 }  \right].
\end{eqnarray}
We now express the trigonometric sine and cosine functions 
in Eq.~(\ref{deltaE4net2}) as exponentials, 
\begin{eqnarray}\label{deltaE4net3}
\fl E_{\mathrm{L}}(R) 
= \frac{-e^4}{72 \pi^3} \sum_{\rho, \sigma}
\sum_{j,\ell}\frac{ \langle \phi_{1S,A} | x^j |\rho\rangle
\langle \rho| x^j |\phi_{1S,A}  \rangle \langle \phi_{1S,B} | x^{\ell} 
|\sigma\rangle \langle \sigma| x^{\ell} |\phi_{1S,B}  \rangle} {E_{A\rho}+ E_{B\sigma}}
\frac{1}{2\mathrm{i}} \left[  \int\limits_0^\infty   
\mathrm{d} k _1\mathrm{e}^{2\mathrm{i}k_1R} \right. \nonumber\\
\fl\times \frac{ k_1^6(E_{A\rho}+ E_{B\sigma}+2 k_1) }{ (E_{B\sigma} +k_1)
(E_{A\rho}+ k_1)}
 \left\{\frac{1}{(k_1R)^2} + \frac{2\mathrm{i}}{(k_1R)^3}-\frac{5}{(k_1R)^4}  
-\frac{6\mathrm{i}}{(k_1R)^5} +\frac{3}{(k_1R)^6}\right\} 
- \int\limits_0^\infty   \mathrm{d} k _1 k_1^6  
\nonumber\\
\fl\times \frac{ (E_{A\rho}+ E_{B\sigma}+2 k_1) 
\mathrm{e}^{-2\mathrm{i}k_1R}}{ (E_{B\sigma}+ k_1)
(E_{A\rho}+ k_1)}\left.  \left\{ \frac{1}{(k_1R)^2}-\frac{2\mathrm{i}}{(k_1R)^3}
-\frac{5}{(k_1R)^4}  +\frac{6\mathrm{i}}{(k_1R)^5} 
+\frac{3}{(k_1R)^6}\right\}\right]\,.
\end{eqnarray} 
Let us  introduce a new variable $\omega$ which has values 
$\omega = -\mathrm{i} \,k_1 $ in the first $k_1$-integral and 
$\omega = \mathrm{i} \,k_1$ in the second 
$k_1$-integral  inside the square bracket $\left[\ \right]$  in  
Eq.~(\ref{deltaE4net3}). 
This amounts to a Wick rotation~\cite{UjVd2017},
which can be carried out without problems because
we are dealing with ground-state atoms,
\begin{eqnarray}\label{deltaE4inU}
\fl \quad E_{\mathrm{L}}(R) = -\frac{e^4}
{72 \pi^3} \int_0^{\infty} \mathrm{d} \omega
\frac{\omega^4\,\mathrm{e}^{ - 2 \omega R}}{R^2} 
\sum_{\rho, j} E_{A\rho}
\frac{ \langle \phi_{1S,A} | x^j |\rho\rangle
\langle \rho| x^j |\phi_{1S,A}
\rangle}{(E_{A\rho}^2 + \omega^2)} 
\nonumber\\
\fl\qquad\times \sum_{\sigma,\ell} E_{B\sigma}\,  
\frac{\langle \phi_{1S,B} | x^{\ell} |\sigma\rangle
 \langle \sigma| x^{\ell} |\phi_{1S,B}  \rangle}%
{ (E_{B\sigma}^2+ \omega^2)}
\left[ 1+ \frac{2}{\omega R} +\frac{5}{(\omega R)^2} 
+ \frac{6}{(\omega R)^3}+\frac{3}{(\omega R)^4} \right] 
\nonumber\\
\fl \qquad = - \frac{1}{32\,\pi^3} \int\limits_0^{\infty}
\mathrm{d}\omega \alpha_{A}(\mathrm{i} \omega) 
\alpha_{B}(\mathrm{i} \omega)
\, \frac{\omega^4  \mathrm{e}^{ - 2 \omega R} }{R^2}
\left[ 1+ \frac{2}{\omega R} +\frac{5}{(\omega R)^2} 
+ \frac{6}{(\omega R)^3}+\frac{3}{(\omega R)^4} \right],
\end{eqnarray} 
where  the quantities $\alpha_{A}(\mathrm{i}\omega)$ and 
$\alpha_{B}(\mathrm{i}\omega)$  are the  dynamic ground-state
polarizabilities of atoms $A$ and  $B$, respectively, 
\numparts
\begin{eqnarray} \label{alpha:A}
\alpha_{A}(\mathrm{i}\omega)&=& \frac{2e^2}{3} \sum_{\rho,j} 
\frac{ E_{A\rho}  }{(E_{A\rho}^2+\omega^2)}
\langle \phi_{1S,A} | x^j |\rho\rangle  \langle \rho|
x^j |\phi_{1S,A}  \rangle \,,
\\ \label{alpha:B}
\alpha_{B}(\mathrm{i}\omega)&=& \frac{2e^2}{3} \sum_{\sigma,\ell}
\frac{ E_{B\sigma}  }{(E_{B\sigma}^2+\omega^2)}
\langle \phi_{1S,B} | x^j |\sigma\rangle  \langle \sigma |
x^j |\phi_{1S,B}  \rangle \,.
\end{eqnarray}
\endnumparts
%
The dynamic  polarizabilities  given in Eqs.~(\ref{alpha:A}) and (\ref{alpha:B}) can be rewritten as 
\numparts
\label{defPPP}
\begin{eqnarray}
\label{polarA}
\alpha_{A}(\mathrm{i}\omega) &=& \frac{e^2}{3}  \sum_{\pm,j} 
\langle \phi_{1S,A} | x^j\frac{ 1}
{H-E_{1S,A} \pm \mathrm{i}\omega} x^j |\phi_{1S,A} \rangle 
= \sum_\pm P_{1S,A}(\pm \mathrm{i} \omega) ,
\\
\label{polarB}
\alpha_{B}(\mathrm{i}\omega) &=& \frac{e^2}{3}  \sum_{\pm,\ell}  
\langle \phi_{1S,B} | x^{\ell}\frac{ 1}
{H-E_{1S,B}\pm \mathrm{i}\omega} x^{\ell} |\phi_{1S,B}  \rangle
= \sum_\pm P_{1S,B}(\pm \mathrm{i} \omega),
\end{eqnarray}
\endnumparts
with an obvious definition of the 
polarizability matrix elements $P$.
For large $\omega$, the polarizabilities show $\omega^{-2}$ behavior. 
The expression for the Casimir-Polder interaction energy
between any two atoms $A$ and $B$
from the six time-ordered diagrams of the ``ladder'' type
(I, III, IV, VII, IX, and X in Fig.~\ref{figg2})
is thus given by 
\begin{eqnarray}\label{eqintomega}
\fl E_{\mathrm{L}}(R)
= -\frac{1}{32 \pi^3} \int\limits_0^{\infty}  
\mathrm{d}\omega \,  \alpha_{A}(\mathrm{i}\omega) 
\alpha_{B}(\mathrm{i}\omega) \frac{\omega^4  
\mathrm{e}^{ - 2\omega R} }{R^2}
\bigg[ 1+  \frac{2}{\omega R}
+\frac{5}{\left(\omega R\right)^2} + \frac{6}{\left(\omega R\right)^3}
+\frac{3}{\left(\omega R\right)^4} \bigg],
\end{eqnarray} 
Here we have used  $e^2=4\pi\alpha$ which holds in the natural units,
and we remark that 
Eq.~(\ref{eqintomega}) is valid for any interatomic separation  
$R$ provided their wave functions do not overlap.

%
%
%
\begin{figure}[t!]
\begin{center}
\includegraphics[width=0.6\linewidth]{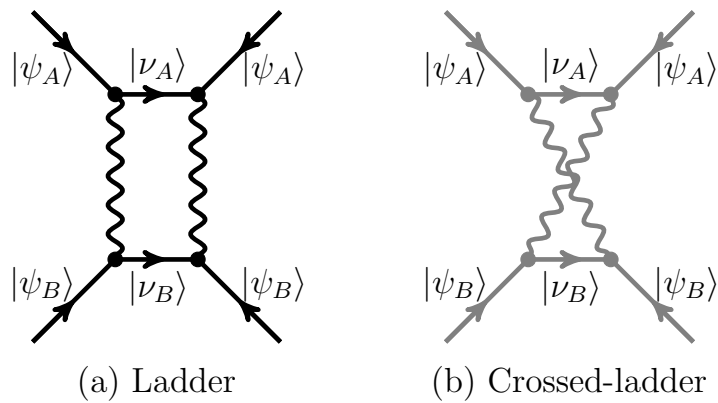}
\caption{\label{figg3}
The ladder (a) and crossed-ladder (b) Feynman diagrams,
shaded in order to show the equivalence to the time-ordered
diagrams in Fig.~\ref{figg1}.
$\left.|\nu_{A}\right>$ and $\left.|\nu_{B}\right>$ are
virtual-states accessible by a dipole transition from
the atomic reference states
$\left.|\psi_{A}\right>$ and $\left.|\psi_{B}\right>$,
respectively. The latter are chosen as the $| 1S \rangle$
states in the calculations reported here}
\end{center}
\end{figure}

Interestingly, the $E_{\mathrm{L}}(R)$ 
is one half of the total Casimir-Polder interaction 
energy between two atoms.
(see Refs.~\cite{UjVd2017}, Chap.~85 of Ref.~\cite{LandauLifshitz4},
or Ref.~\cite{CAThesis2017}).
The other half to the Casimir-Polder interaction,
denoted here as $E_{\mathrm{C}}(R)$, 
where the subscript ``C'' stands for cross,
comes from the remaining six-time-ordered diagrams 
in which photon lines cross, namely,
II, V, VI, VIII, XI, and XII in Fig.~\ref{figg1}.
One can perform a separate evaluation of these
crossed diagrams, along the same ideas as discussed
above (in particular, the integration contours 
in Fig.~\ref{figg2} are useful).
Skipping further details, it is useful to point out that
the contribution of the six time-ordered 
diagrams with crossed photon lines is just the same 
as the one from the ladder diagrams, i.e., that 
$E_{\mathrm{C}}(R) = E_{\mathrm{L}}(R)$.

%
%
%
\section{Covariant Formalism: Matching the Scattering Amplitude}
\label{sec3}

We briefly recall the formalism used in Ref.~\cite{UjVd2017},
in order to identify the contribution of the 
crossed and ladder diagrams to the Casimir--Polder 
interaction energy.
To the  fourth-order, the  contribution to  the scattering operator, 
$\widehat{S}$, is given by  the following expression 
(see Eq.~(5) of Ref~\cite{UjVd2017})
\begin{eqnarray}
\label{TTT}
\widehat{S}^{(4)}= \frac{1}{24} \int_{-\infty}^{\infty} \mathrm{d}t_1 
\int_{-\infty}^{\infty} \mathrm{d}t_2 \int_{-\infty}^{\infty} \mathrm{d}t_3 
\int_{-\infty}^{\infty} \mathrm{d}t_4\,
\widehat{T} [V(t_1) V(t_2) V(t_3)V(t_4)]\,,
\end{eqnarray}
where $\widehat{T}$ denotes the time ordering operator.
In the dipole approximation, the interaction Hamiltonian
$V(t) = \widehat{H}_{AB}$ (see Eq.~(\ref{HAB})) can be conveniently
expressed as 
\begin{eqnarray}\label{approxVtef}
V(t)\approx &-  \vec{d}_A\cdot \vec{E}(\vec{R}_A,t) 
-   \vec{d}_B\cdot \vec{E}(\vec{R}_B,t)\,,
\end{eqnarray}
where $\vec{d}_i= e\, \vec{r}_i$  is the  electric dipole operator 
for  atom $i$ whose nucleus is at  $\vec{R}_i$,
and this time, we explicitly indicate the time-dependence
of the interaction Hamiltonian,
employing interaction-picture field operators.  
Assuming that the unperturbed state of the system contains 
atoms on the state $|\psi\rangle= |\psi_A,\psi_B\rangle$ and 
the electromagnetic field in the vacuum state $|0\rangle$, 
the fourth-order forward-scattering $S$-matrix 
element is given by
\begin{eqnarray}\label{s4opertarf}
\langle S^{(4)} \rangle =& \langle \psi|\langle 0| \widehat{S}^{(4)}  
|0\rangle |\psi\rangle\,.
\end{eqnarray}
The time ordering of the electric-field operators
in Eq.~(\ref{TTT}) leads to the photon propagators,
while the four types of contributions which arise due to 
time orderings of electric dipole moment operators 
in the interactions $V(t_i)$ are given in Eq.~(6) of Ref.~\cite{UjVd2017},
which read as follows:
\numparts
\label{C-terms}
\begin{eqnarray}
C_1 &\equiv & \; 
\langle \psi_A | \widehat{T}_d \; d_{Ai}(t_1) \, d_{Ak}(t_3) | \psi_A \rangle 
\;
\langle \psi_B | \widehat{T}_d \; d_{Bj}(t_2) \, d_{B\ell}(t_4) ) | \psi_B \rangle \,,
\\
C_2& \equiv & \; 
\langle \psi_A | \widehat{T}_d \; d_{Ai}(t_1) \, d_{A\ell}(t_4) | \psi_A \rangle 
\;
\langle \psi_B | \widehat{T}_d \; d_{Bj}(t_2) \, d_{Bk}(t_3) | \psi_B \rangle \,,
\\
C_3 &\equiv & \; 
\langle \psi_B | \widehat{T}_d \; d_{Bi}(t_1) \, d_{B\ell}(t_4) | \psi_B \rangle 
\;
\langle \psi_A | \widehat{T}_d \; d_{Aj}(t_2) \, d_{Ak}(t_3) | \psi_A \rangle \,,
\\
C_4&\equiv & \; 
\langle \psi_B | \widehat{T}_d \; d_{Bi}(t_1) \, d_{Bk}(t_3) | \psi_B \rangle 
\; 
\langle \psi_A | \widehat{T}_d \; d_{Aj}(t_2) \, d_{A\ell}(t_4) | \psi_A \rangle \,,
\end{eqnarray}
\endnumparts
where $\widehat{T}_d$ is the time ordering 
operator for the  dipole moments.
The  graph (a) of~Fig.~\ref{figg3} 
represents the sum of the contributions $C_1$ and $C_3$, while 
the other two contributions, viz.~$C_2$ and $C_4$ come from the 
graph (b).  All terms given in
Eq.~(\ref{C-terms}) are multiplied by two photon 
propagators of the index structure ``$12$'' and ``$34$'',
respectively, and, hence, give identical contributions to the 
$S$ matrix element, 
as explained in detail in Ref.~\cite{UjVd2017}.
Consequently, the  
ladder $\langle S^{(4)} \rangle_{\mathrm{L}} $ and 
the crossed-ladder $\langle S^{(4)} \rangle_{\mathrm{C}} $
contributions to the scattering matrix element can be written as  
\begin{eqnarray}
\label{S4LS4CL}
\fl\langle S^{(4)} \rangle_{\mathrm{L}}=\langle S^{(4)} \rangle_{\mathrm{C}} 
&=&  \frac{1}{4} \int_{-\infty}^{\infty} \mathrm{d}t_1  
\int_{-\infty}^{\infty} \mathrm{d}t_2 
\int_{-\infty}^{\infty} \mathrm{d}t_3  
\int_{-\infty}^{\infty}\mathrm{d}t_4 
  \langle 0|\widehat{T}_E \Big[{E}_i(\vec{R}_A,t_1)\, 
 {E}_j(\vec{R}_B,t_2)\Big]|0\rangle\nonumber\\
\fl&\times&\langle 0|\widehat{T}_E \Big[{E}_k(\vec{R}_A,t_3)\, 
{E}_\ell(\vec{R}_B,t_4)\Big]|0\rangle
\langle \psi_A|\widehat{T}_d\Big[ d_{Ai}(t_1)\, 
d_{Ak}(t_3)\Big] |\psi_A\rangle \nonumber\\
\fl &\times&\langle \psi_B|\widehat{T}_d\Big[ d_{Bj}(t_2)\,
d_{B\ell}(t_4)\Big] |\psi_B\rangle\,.
\end{eqnarray}
At this point, we could stop the 
calculation and argue that, since the $S$ matrix 
elements generated by the crossed and ladder diagrams
are the same, the effective Hamiltonians and 
energy shifts corresponding to the diagrams
also must be the same, proving consistency 
with the results of Sec.~\ref{sec2}. However, we 
carry through the derivation for completeness.
We recall that $\widehat{T}_E$ is the time ordering operator for the  
electric field operators. According to Eqs.~(18) and (21) 
of Ref.~\cite{UjVd2017}, one may carry out the $t$-integrals of 
Eq.~(\ref{S4LS4CL}), which finally gives
\begin{eqnarray}\label{s4avgfinalexpagainF}
\langle S^{(4)} \rangle_{\mathrm{L}}=  \frac{T}{4}  
\int_{-\infty}^\infty 
\frac{\mathrm{d}\omega}{2\pi}\, \omega^4\,  
D_{ij}( \omega,\vec{R}) 
 D_{k\ell}( \omega,\vec{R})   
\alpha_{A,ik}(\omega)  \alpha_{B,j\ell}(\omega)\,,
\end{eqnarray}
where $T=\int_{t_i}^{t_f} \mathrm{d} t= t_f-t_i$ 
denotes  the total interval of time  in which the transition occurs.
The photon propagator, or, merely,
as explained in Ref.~\cite{UjVd2017}, 
electric-field propagator, $D_{ij}(\omega, \vec{R})$, 
can be expressed in terms of the 
tensor structures $\alpha_{ij} $ and $ \beta_{ij} $,
\begin{eqnarray}\label{omegaRDij}
D_{ij}(\omega, \vec{R}) 
=&  \frac{ \mathrm{e}^{\mathrm{i} |\omega|R} }
{ 4\pi }\left[ \mathcal{\alpha}_{ij} - \beta_{ij} 
\left( \frac{\mathrm{i}}{|\omega | R} -\frac{1}{\omega^2 R^2}\right) 
\right]\,,
\end{eqnarray}
where 
\begin{eqnarray}\label{alphabeta}
\alpha_{ij} =  \delta_{ij} - \frac{R_i R_j}{R^2}, \quad  \text{and} \quad 
 \beta_{ij} =  \delta_{ij} -3 \frac{R_i R_j}{R^2}\,. 
\end{eqnarray} 
The dynamic polarizability  
$\alpha_{A,ik}(\omega)$ in Eq.~(\ref{s4avgfinalexpagainF})
is given as
\begin{eqnarray}\label{omegaRAlphaij}
\fl\alpha_{A,ik}(\omega)=  \sum_{\nu_A} \left( \frac{\langle\psi_A| 
d_{Ai}|\nu_A\rangle \, \langle \nu_A |d_{Aj}|\psi_A\rangle}
{ E_{\nu,A}- \omega - \mathrm{i}\epsilon}
+ \frac{\langle\psi_A| 
d_{Ai}|\nu_A\rangle \, \langle \nu_A |d_{Aj}|\psi_A\rangle}
{ E_{\nu,A} + \omega - \mathrm{i}\epsilon}\right) \,.
\end{eqnarray}
The matching relation for the diagonal element of the 
effective Hamiltonian $H_{\rm eff}$ 
(``quasipotential'') derived from
an $S$ matrix element is (see Eq.~(3) of Ref.~\cite{UjVd2017})
\begin{eqnarray}
\langle S^{(4)} \rangle = -\mathrm{i}\, T
\langle \psi | H_{\rm eff} |\psi\rangle \,,
\end{eqnarray}
so that, for the contribution of the ladder Feynman diagram of 
$\langle \psi | H_{\rm eff} |\psi\rangle = E_{\rm L}(R)$,
one has in view of Eq.~(\ref{s4avgfinalexpagainF}) 
\begin{eqnarray}
\label{deltaEL}
E_{\mathrm{L}}(R) = \frac{\mathrm{i}}{2} \int_0^{\infty} 
\frac{\mathrm{d}\omega}{2\pi} 
\omega^4 \, D_{ij}(\omega, \vec{R}) \, D_{k\ell} (\omega, \vec{R})
\alpha_{A,ik}(\omega)\,  \alpha_{B,j\ell}(\omega)\,.
\end{eqnarray}
 For a reference $1S$ state,
one has $\alpha_{A,ik}(\omega) = (\delta^{ik}/3) \,
\alpha_{A}(\omega)$.
Under a Wick-rotation $\omega \to \mathrm{i} \, \omega$,
Eq.~(\ref{deltaEL}) reads as
\begin{eqnarray}
\label{ELFinal}
E_{\mathrm{L}}(R)
&&= -\frac{1}{32 \pi^3} \int_0^{\infty}  
\mathrm{d}\omega\, \alpha_{A}(\mathrm{i}\omega) \,
\alpha_{B}(\mathrm{i}\omega)\, \frac{\omega^4  
\mathrm{e}^{ - 2\omega R} }{R^2}\nonumber\\
&&\times  \left[ 1+  \frac{2}{\omega R}
+\frac{5}{\left(\omega R\right)^2} + \frac{6}{\left(\omega R\right)^3}
+\frac{3}{\left(\omega R\right)^4} \right]\,.
\end{eqnarray} 
Note that, $E_{\mathrm{L}}(R)$ is 
half of the total interaction energy,
confirming the consistency with the 
result reported in Sec.~\ref{sec2},
which implies that the ladder-type diagrams
contribute exactly half of the 
Casimir--Polder interaction.

%
%
\section{Radiative Corrections}
\label{sec4}

Relativistic corrections to the leading-order expression 
\begin{eqnarray}
\label{leadingexpr}
E(R) &=& E_{\rm L}(R) + E_{\rm C}(R) =
-\frac{1}{16 \, \pi^3} \int_0^{\infty}
\mathrm{d}\omega\, \alpha_A(\mathrm{i}\omega) \,
\alpha_B(\mathrm{i}\omega)\, \frac{\omega^4
\mathrm{e}^{ - 2\omega R} }{R^2}\nonumber\\
&&\times  \left[ 1+  \frac{2}{\omega R}
+\frac{5}{\left(\omega R\right)^2} + \frac{6}{\left(\omega R\right)^3}
+\frac{3}{\left(\omega R\right)^4} \right] .
\end{eqnarray}
involve 
corrections to the atomic Hamiltonian, 
to the energy, to the wave function, and to the transition 
current~\cite{Pachucki2005}. 
In units with $\hbar = c = \epsilon_0 = 1$,
which are used throughout this article,
the Bohr radius is $a_0 = (\alpha m)^{-1}$,
and the interatomic distance,
expressed in atomic units, is 
\begin{eqnarray}
\label{defrho}
\rho = \frac{R}{a_0} = \alpha m R \, .
\end{eqnarray}
One can write (see Ref.~\cite{Pachucki2005})
a systematic expansion of the interaction energy,
which clarifies the relevant orders of the expansion 
in powers of the fine-structure constant~$\alpha$.
In the non-retardation regime, one encounters the following terms
[see Eq.~(29) of Ref.~\cite{Pachucki2005}]
\begin{eqnarray}
\label{ER}
E(\alpha, mR) = \; E_{\rm free}(\alpha) -
\sum_{i,j} m \alpha^i \frac{ C_j^{(i)} }{ (m \alpha R )^j } 
=\; E_{\rm free}(\alpha) -
\sum_{i,j} m \alpha^i \frac{ C_j^{(i)} }{ \rho^j } \,,
\end{eqnarray}
where $E_{\rm free}(\alpha)$ refers to the $\alpha$-expansion
of the sum of the free (non-interacting) energies 
of the two atoms. 
[For a conjectured necessary generalization of this 
expansion, see Eq.~(\ref{expansion}) below.]

The leading term from the 
Casimir-Polder interaction is proportional to $C_6^{(2)}$,
and equal to the van der Waals energy.
Here, we recall that in our units, the 
Hartree energy is expressed as $E_h = \alpha^2 m $.
In the non-retardation regime,
the quadrupole term gives a correction proportional 
to $C_8^{(2)}$ (see Eq.~(33) of Ref.~\cite{Pachucki2005}).
Surprisingly, this term is not suppressed by 
a factor of $\alpha$, but by a higher power of the 
scaled interatomic distance $\rho$
(eighth power instead of sixth).
The relativistic corrections to the 
Hamiltonian, energy and wave function,
together with the dipole-octupole mixing term and 
the relativistic corrections to the current,
give the term $C_6^{(4)}$, which is 
still proportional to $1/\rho^6$, but has a 
prefactor $m \alpha^4$ instead of $m \alpha^2$, 
and is thus suppressed by two powers of $\alpha$.
We find that the radiative correction 
to the Casimir-Polder interaction contributes, 
in the non-retardation regime, 
to the coefficient $C_6^{(5,1)}$,
with a prefactor proportional to 
$\alpha^5 \, m \, \ln(\alpha^{-2})$.
(The single power of the logarithm is denoted 
here by the second upper index of the coefficient,
inspired by a commonly accepted notation
adopted in Lamb shift calculations~\cite{UdKp1996}.)

In order to obtain the leading radiative
correction, we use the ``effective 
radiative Lamb shift potential'' (see Ref.~\cite{Udj2003}),
denoted as $\delta V_{\rm Lamb}$,
\begin{eqnarray}
\label{VLamb}
\delta V_{\rm rad} = \frac{4 \alpha}{3 \pi} \,
[\pi \alpha] \,\ln[\alpha^{-2}] \, \frac{\delta^{(3)}(\vec{r})}{m^2}
= \frac{4 \alpha}{3 \pi} \, \ln[\alpha^{-2}] \, \delta V\,,
\end{eqnarray}
where $\delta V$ is a ``standard potential'' whose 
expectation value, on a hydrogenic state,
has particularly simple prefactors,
\begin{eqnarray}
\label{standard}
\delta V 
= \frac{\pi \alpha}{m^2} \, \delta^{(3)}(\vec{r})
= \pi \alpha^4 m \, \delta^{(3)}\left( \frac{\vec{r}}{a_0} \right)\,,
\qquad
\left< nS \left| \delta V \right| nS \right> = \frac{(\alpha m)^3}{n^3} \,.
\end{eqnarray}
We recall that only $S$ states are nonvanishing at the origin.
We then perturb the Hamiltonian, energy, and reference state,
by the radiative Lamb shift potential,
in both atoms $A$ and $B$.
We will study the corresponding radiative shift
for two hydrogen atoms, which are both in their ground state.
The modification of the total Casimir-Polder interaction 
can be written as 
\begin{eqnarray}
\label{DeltaEcpInd}
\delta  E(R)
&=& -\frac{2}{ \pi (4 \pi)^2} \int_0^{\infty}  
\mathrm{d}\omega \,  \alpha_A(\mathrm{i}\omega) \,
\delta\alpha_B(\mathrm{i}\omega)\,
\frac{\omega^4  
\mathrm{e}^{ - 2\omega R} }{R^2}\nonumber\\
& & \times \left[ 1+ \frac{2}{\omega R} 
+\frac{5}{\left(\omega R\right)^2 }
+ \frac{6}{\left(\omega R\right)^3}
+\frac{3}{\left(\omega R\right)^4} \right] \,,
\end{eqnarray} 
where $\delta\alpha_B$ is the 
perturbation of the polarizability due to the 
``standard potential''~(\ref{standard}),
\begin{eqnarray}
\delta\alpha_B(\mathrm{i}\omega)=\sum_{\pm} 
\delta P_{1S}(\pm\mathrm{i}\omega)\,.
\end{eqnarray}
Here, $\delta P_{1S}$ is the $\delta V$-induced 
perturbation of the polarizability matrix element
defined in Eq.~(\ref{defPPP}), for atom $B$.
The perturbed $P$-matrix $\delta P_{1S}$ element 
has three contributions, namely, corrections to the  Hamiltonian 
of the propagator denominator, the energy,
and the wave function. Explicitly,
\begin{eqnarray}
\delta P_{1S}(\mathrm{i}\omega)=
\delta P^{(H)}_{1S}(\mathrm{i}\omega)
+\delta P^{(E)}_{1S}(\mathrm{i}\omega)
+\delta P^{(\phi)}_{1S}(\mathrm{i}\omega)\,,
\end{eqnarray}
The correction arising from the Hamiltonian reads 
\begin{eqnarray}
\delta P^{(H)}_{1S}(\mathrm{i}\omega)
=-\frac{e^2}{3}\left< 1S\left|x^i 
\frac{1}{H-E_{1S}+\mathrm{i}\omega}
\; \delta V\; 
\frac{1}{H-E_{1S}-\mathrm{i}\omega} x^i\right| 1S\right>\,,
\end{eqnarray}
which is zero as the matrix element of the 
Dirac-$\delta$ between any two 
virtual $P$ states vanishes. 
The contribution to the Casimir-Polder interaction 
from the correction to the energy  is given by 
\begin{eqnarray}
\delta P^{(E)}_{1S}(\mathrm{i}\omega)
=-\frac{\partial}{\partial\,\omega}
P_{1S}(\mathrm{i}\omega)\left< 1S\left|
\delta V
\right| 1S\right>
= -\alpha^4m \, 
\frac{\partial}{\partial\,\omega}
P_{1S}(\mathrm{i}\omega)\,,
\end{eqnarray}
The modification of the $P$-matrix element due to the 
wave function correction, to the first order, is given by
\begin{eqnarray}
\delta P^{(\phi)}_{1S}(\mathrm{i}\omega)
=\frac{2}{3} e^2
\left< 1S\left|x^i 
\frac{1}{H-E_{1S}+\mathrm{i}\omega}
 x^i\right| \delta(1S)\right>\,,
\end{eqnarray}
where the perturbed $1S$-state wave function is
\begin{eqnarray}
\left| \delta(1S)\right> =
\frac{1}{(E_{1S} - H)'} \, \delta V \, | 1S \rangle \,.
\end{eqnarray}
In coordinate space, one has
\begin{eqnarray}
\label{deltapsi1S}
\fl \delta \Psi_{1S}(\vec{r})=\frac{1}{\sqrt{4\pi}} \delta R_{10}(r)
=2 \alpha^2 
\frac{\mathrm{e}^{-r/a_0}}{3\pi\sqrt{\pi a_0}}
\left[-\frac{1}{r}-\frac{1}{a_0}\left( 5-2\gamma_{_{E}}-
2\ln\left(\frac{r}{a_0}\right)\right)
+\frac{2r}{a_0^2}\right]\,,
\end{eqnarray}
where $\gamma_{_{E}}=0.577\,2157$  is the
Euler--Mascheroni constant.
The result~(\ref{deltapsi1S}) is in agreement with 
Eq.~(23) of Ref.~\cite{CoDa1961}.

%
%
\subsection{Short Range}

In the short-range regime, i.e., $1/(\alpha m)\ll R\ll 1/(\alpha^2m)$\,,
the radiative correction to the 
interaction energy takes the form 
\begin{eqnarray}
\delta E_{\rm rad}(\alpha, m R)&=&
-\frac{6}{\pi (4 \pi)^2 R^6}\int_0^\infty \mathrm{d}\omega\,
\alpha_{1S}(\mathrm{i}\omega)\,
\delta \alpha_{1S}(\mathrm{i}\omega)\,
\frac{4\alpha}{3\pi} \,\ln\left(\alpha^{-2}\right)
\nonumber\\
&=&  -\frac{4}{3\pi} \,
\alpha^3 m \,\ln\left(\alpha^{-2}\right)
\frac{\delta X^{\rm dl}}{\left(\alpha mR\right)^6} \,,
\end{eqnarray}
where the delta-perturbed van der Waals  $\delta X^{\rm dl}$
coefficient (``dl'' stands for dimensionless, i.e., expressed in atomic 
units) is given by 
\begin{eqnarray}\label{deltaC6dl}
\delta X^{\rm dl}&=& \frac{6}{\pi}\int_0^\infty \mathrm{d}\omega\,
\alpha^{\mathrm{dl}}_{1S}(\mathrm{i}\omega)\,
\delta \alpha^{\mathrm{dl}}_{1S}(\mathrm{i}\omega)
\nonumber\\
&=& \frac{6}{\pi}\int_0^\infty \mathrm{d}\omega\,
\alpha_{1S}^\mathrm{dl}(\mathrm{i}\omega)\,
\left( \delta \alpha_{1S}^{(E,\mathrm{dl})}(\mathrm{i}\omega)+
\delta \alpha_{1S}^{(\phi,\mathrm{dl})}(\mathrm{i}\omega)\right)\,.
\end{eqnarray}
Here $E$, $\phi$, and dl 
stand for the energy correction, the wave function part,  and the 
dimensionless quantity respectively.
One can use convergence acceleration techniques as discussed in 
Refs.~\cite{Udj.Con.Acc.Tech1999,UdjThesis2003}  in addition to 
other numerical methods presented in Ref.~\cite{UdKp1996}
and evaluate the integral~(\ref{deltaC6dl}) numerically,
which yields
\begin{eqnarray}
\delta X^{\mathrm{dl}}=69.371\,0888\, \alpha^2\,.
\end{eqnarray}
The radiative correction to the 
interaction energy can be expressed as  
\begin{eqnarray}\label{delErc}
\delta E_{\rm rad}(\alpha, m R)
=- \alpha^5 m \, 
\frac{\delta C_6^{(5)}}{\left(\alpha mR\right)^6}\,,
\end{eqnarray}
where 
\begin{eqnarray}
\delta C_6^{(5)}= 29.442\,0042 \, \ln( \alpha^{-2} )
\end{eqnarray}
is a large coefficient, which, in addition,
also contains a logarithm of the fine-structure constant.
The large magnitude of the logarithmic coefficient multiplying the
radiative correction,
which amounts to an approximate 
numerical value of $30 \times \ln( 137^2 ) \approx 300$,
compensates the additional power of $\alpha$ in comparison
to the relativistic corrections considered in 
Ref.~\cite{Pachucki2005}; this 
implies that the effect is of the same order-of-magnitude
as the relativistic corrections considered in 
Ref.~\cite{Pachucki2005} and should be included
in any precise theory of the interatomic interaction.
In a wider context, 
the emergence of logarithmic terms
in an accurate treatment of the interatomic interaction,
in both the interatomic distance as well 
as the fine-structure constant, is discussed in the Appendix.
On the other hand, in the short-range 
regime, the interaction energy $E(\alpha, mR)$
is given by~\cite{OurPRA1from2017}
\begin{eqnarray}\label{EalphamR}
E(\alpha, mR) =
-\frac{3\,\alpha^2}{\pi e^4 R^6}\int_0^\infty \mathrm{d}\omega\,
\alpha_{1S}(\mathrm{i}\omega)\,
\alpha_{1S}(\mathrm{i}\omega)
 =-\alpha^2 m \frac{C_6^{(2)}}{\left(\alpha mR\right)^6}\,.
\end{eqnarray}
where $C_6^{(2)} = 6.499\,0267$.
(In obtaining numerical results,
we treat the hydrogen atoms in the non-recoil limit,
i.e., in the limit of an infinite mass of the nucleus.)
Comparing $\delta E_{\rm rad}(\alpha, mR)$ of 
Eq.~(\ref{delErc}) to  $E(\alpha, mR)$
as given in Eq.~(\ref{EalphamR}), 
one can conclude that  the correction to the Casimir-Polder interaction
due to the leading  radiative correction is  
of relative order $\alpha^3\,\ln\left(\alpha^{-2}\right)$.

%
%
\subsection{Long Range}
 
In the long-range limit, i.e., $R\gg 1/(\alpha^2m)$\,, however, 
the dynamic polarizability of the 
ground state  can be 
approximated by its static polarizability. Consequently, the Casimir-Polder interaction
and the radiative correction to the Casimir-Polder interaction read
\begin{eqnarray}
E(\alpha, mR) &=& -\alpha_A(0) \, 
\alpha_B(0)\, f(R)\,,\\
\delta E_{\rm rad}(\alpha, mR) &=&- 2\,\alpha_A(0)  \,
\delta\alpha_B(0)
\frac{4\alpha}{3\pi}\ln\left(\alpha^{-2}\right) f(R),
\end{eqnarray}
where the function $f(R)$ is an integral over 
the angular frequency $\omega$,
\begin{eqnarray}
\fl \;f(R)=\frac{1}{16 \pi^3 R^2}\int_0^{\infty}  
\mathrm{d}\omega \,\omega^4   
\mathrm{e}^{ - 2\omega R}
 \left[ 1+ \frac{2}{\omega R}
+\frac{5}{\left(\omega R\right)^2}\right.
\left. + \frac{6}{\left(\omega R\right)^3}
 + \frac{3}{\left(\omega R\right)^4}\right]
=\frac{23}{(4\pi)^3 R^7}.
\end{eqnarray}
The ground state static polarizability $\alpha_A(0)$ 
of atom $A$, in the case of hydrogen, is given by 
\begin{eqnarray}
\alpha_A(0)=\frac{ 9\, e^2}{2\,\alpha^4 m^3}\,.
\end{eqnarray}
The  $\delta V$-perturbed
ground state static polarizability $\delta\alpha_B(0) = 
\delta\alpha_{1S}(0)$ is the sum
\begin{eqnarray}
\delta\alpha_{1S}(0)=\delta\alpha^{(E)}_{1S}(0)+\delta\alpha^{(\phi)}_{1S}(0)
=\frac{167\,e^2}{46\, \alpha^2 \, m^3}\,,
\end{eqnarray}
where
\begin{eqnarray}
\delta\alpha^{(E)}_{1S}(0)=\frac{43\,e^2}{23\, \alpha^2 \, m^3}\,,
\qquad
\delta\alpha^{(\phi)}_{1S}(0)=\frac{81\,e^2}{46\, \alpha^2 \, m^3}\,,
\end{eqnarray}
are, respectively,  the energy and the wave function parts of  delta perturbed 
ground state static polarizability.
As a result,  we have, in natural units,
\begin{eqnarray}\label{ULRCpE7}
E(\alpha, mR)
=-\alpha^8m\, \frac{1863}{16\,\pi}\,  \frac{1}{\left(m\alpha^2 R\right)^7}\,,
\\
\label{ULRdeltaCE7}
\delta E_{\rm rad}(\alpha, mR)
= -\alpha^8m\, \frac{501}{2\,\pi^2}\,  
\frac{\alpha^3 \,
\ln\left(\alpha^{-2}\right)}{\left(m\alpha^2 R\right)^7} \,.
\end{eqnarray}
It is evident from Eqs.~(\ref{ULRCpE7}) and (\ref{ULRdeltaCE7}) that,
in the  long-range, the  Casimir-Polder interaction 
and the perturbed  Casimir-Polder interaction 
vary as inverse seventh 
powers of the interatomic distance, and the 
leading-order radiative correction 
to the Casimir-Polder interaction is 
of relative oder $\alpha^3\,\ln\left(\alpha^{-2}\right)$.

%
%
\section{Conclusions}
\label{sec5}

We have analyzed the Casimir-Polder interaction between 
two neutral  hydrogen atoms in the ground state.
This process entails the exchange of two virtual photons.
The topologically distinct 12 time-ordered diagrams  
are grouped into two equal half on the basis of 
the presence of crossing in the photon-lines (see Sec.~\ref{sec2}).  
The contribution $E_{\mathrm{L}}$ of the six ``ladder'' diagrams, 
in which the photon lines do not cross,
is seen to be equal to the contribution $E_{\mathrm{C}}$
of the six diagrams with crossing photon lines.

Within the framework of covariant form of  
Quantum Electrodynamics, all of these twelve time-ordered
diagrams can be replaced by just 
two Feynman diagrams (Sec.~\ref{sec3}). 
The contribution of the ladder Feynman diagram is seen to 
equal the contribution of the six ``ladder'' diagrams
(without crossed photon lines)
in the time-ordered formalism.
In addition to this, 
the time-ordering formalism and the covariant 
formalism yield identical results
for the total Casimir-Polder interaction. 

In Sec.~\ref{sec4}, we discuss a systematic 
expansion of the Casimir--Polder interaction
energy in powers of the interatomic distance, and of the 
fine-structure constant.
In the sense of Eq.~(\ref{ER}),
we find that the radiative correction 
to the Casimir-Polder interaction contributes, 
in the non-retardation regime, 
to the coefficient $C_6^{(5,1)}$,
with a logarithmic factor.
Specifically, it is proportional to 
$\alpha^5 \, m \, \ln(\alpha^{-2})/\rho^6$,
where $\rho$ is the interatomic distance,
measured in atomic units (see Eq.~(\ref{defrho})). 
(The one power of the logarithm is denoted 
here by the second upper index of the coefficient.) 
As a consequence, the  radiative correction is 
of relative order $\alpha^3\,\ln\left(\alpha^{-2}\right)$.

Our detailed calculation in the time-ordered
formalism, as outlined in Sec.~\ref{sec2},
crucially depends on the correctness of the 
principal-value prescription used in the 
evaluation of the $k_1$ and $k_2$ integrals
given in Eq.~(\ref{Integralk2witheta}).
This treatment is restricted in validity to the 
ground-state interaction, where no additional 
poles due to virtual resonant transitions
to energetically lower virtual states 
are available~\cite{UjVd2017}.
As much as our calculation shows the 
mutual consistency of the time-ordered, and the 
Feynman diagram treatment (the latter profits from the 
matching of the scattering amplitude),
it also highlights the limitations of the 
time-ordered formalism, which avoids making 
concrete statements regarding the correct
placement of the poles of the atomic Green function.

%
%
\section*{Acknowledgments}

This research has been Supported by  the National
Science Foundation (grant PHY-1710856).

\myapp


%
%
\section{Appendix}

We recall, for convenience,
the most general form of the interatomic  Casimir-Polder  interaction 
in term of the dynamic polarizabilities from Eq.~(\ref{leadingexpr}),
where we introduce the variable $x = 2 \omega R$, 
\begin{eqnarray}
\label{x_expr}
\hspace{-2cm}
E(R) = -\int\limits_0^{\infty}
\frac{\mathrm{d} x \, \mathrm{e}^{-x}
\left( 48 + 48 \,x + 20 x^2 + 4 x^3 + x^4 \right) }{512 \pi^3 R^7} \, 
\alpha_A\left(\frac{\mathrm{i} x}{2 R} \right) \,
\alpha_B\left(\frac{\mathrm{i} x}{2 R} \right) \,.
\end{eqnarray}
We would like to find an expansion of this expression
in the range $R > a_0 = 1/(\alpha m)$, but not 
necessarily $R \gg a_0$.
One may use the expression in terms
of oscillator strengths $f_{nA}$ for 
the dynamic polarizability $\alpha_A({\rm i} \omega)$,
\begin{eqnarray}
\label{Eq:pol}
\alpha_A({\rm i} \omega) =
\sum_{n} \frac{f_{nA}}{ \omega_{nA}^2 + \omega^2} \,,
\end{eqnarray}
and analogously for atom $B$.
For a hydrogen atom in the ground states, the 
oscillator strength $f_{nA}$ is given as 
\begin{eqnarray}
f_{nA}=\frac{2 \, e^2 }{3} \, E_{nA} 
\left| \left<\, 0\left|\, \vec{r}_A\,\right| \,n\, \right>\right|^2 \,,
\end{eqnarray}
and otherwise one has to sum over the coordinates of the atomic 
electrons.

In the interatomic distance range relevant to the 
van der Waals interaction, we seek to find the coefficients 
in the expansion [see Eq.~(\ref{ER}) here and 
Eq.~(29) of Ref.~\cite{Pachucki2005}]
\begin{eqnarray}
\label{expansion}
E(\alpha, mR)= -\sum_{i,j} m \alpha^i \,
\frac{C^{(i)}_j}{(\alpha m R)^j} \,,
\end{eqnarray}
where we ignore the free atomic energy.
We here conjecture that the functional form 
given in Eq.~(\ref{expansion}) should be augmented 
by logarithmic terms,
\begin{eqnarray}
\label{expansion2}
E(\alpha, mR)= 
-\sum_{i,j,k} m \alpha^i \,
\frac{C^{(i,k)}_{j} \, \ln^k(\alpha m R)}{(\alpha m R)^j} \,.
\end{eqnarray}
The $C^{(i)}_j$ coefficients are a special case of the 
$C^{(i,k)}_{j}$ for $k=0$.
In order to bring the expressions for the 
coefficients into a convenient form,
one scales variables according to 
\begin{eqnarray}
\hspace{-1cm}
\vec r_A &=& a_0 \, \vec \rho_A \,,
\qquad 
\vec r_B = a_0 \, \vec \rho_B \,,
\qquad 
\vec p_A = \frac{1}{a_0} \, \vec P_A \,,
\qquad 
\vec p_B = \frac{1}{a_0} \, \vec P_B \,,
\\[0.1133ex]
\hspace{-1cm}
H_A &=& E_h \, \calH_A \,,
\qquad
H_B = E_h \, \calH_B \,,
\qquad
E_A = E_h \, \calE_A \,,
\qquad
E_B = E_h \, \calE_B \,,
\end{eqnarray}
where $a_0 = 1/(\alpha m)$ is the Bohr radius,
and $E_h = \alpha^2 m$ is the Hartree energy.
Furthermore,
\begin{eqnarray}
\rho = \alpha m R = R/a_0
\end{eqnarray}
is the interatomic distance, expressed 
in Bohr radii.
The advantage of the scaled variables
$\vec \rho_{A,B}$, 
$\vec P_{A,B}$, 
$\calH_{A,B}$, and
$\calE_{A,B}$ is that they assume 
numerical values and expectation values of 
order unity, for atomic reference states and 
transition matrix elements.
Alternatively, one might say that the 
scaled variables are expressed in ``atomic units''.

We confirm the results given in Eqs.~(30)--(32)
of Ref.~\cite{Pachucki2005},
\begin{eqnarray}
C^{(2,0)}_{6} &=&
\frac23 \, \left< \rho^i_A \rho^j_B
\frac{1}{\calH_A + \calH_B - \calE_A - \calE_B}
\rho^i_A \rho^j_B \right> \,,
\\[0.1133ex]
C^{(4,0)}_{4} &=&
\frac29 \, \left< \rho^i_A \rho^j_B
\frac{1}{\calH_A + \calH_B - \calE_A - \calE_B}
P^i_A P^j_B \right> \,,
\\[0.1133ex]
C^{(5,0)}_{3} &=&
\frac{7}{6 \pi} \, N_A \, N_B \,.
\end{eqnarray}
Here, $N_A$ and $N_B$ are the number of electrons
in atoms $A$ and $B$.
Furthermore, we find the following
representation for the higher-order coefficient 
$C^{(6,0)}_{2}$ emerging
from Eq.~(\ref{x_expr}),
\begin{eqnarray}
C^{(6,0)}_{2} &=&
- \frac13 \, \left( 
N_A \, \langle \vec P^2_B \rangle +
N_B \, \langle \vec P^2_A \rangle 
\right) 
\nonumber\\[0.1133ex]
& & + \frac29 \, \left< P^i_A P^j_B
\frac{1}{\calH_A + \calH_B - \calE_A - \calE_B}
P^i_A P^j_B \right> \,.
\end{eqnarray}
In the seventh order in $\alpha$, a logarithmic term
is obtained, which is proportional to $\rho^{-1}$.
The mechanism behind the generation of the 
logarithm is that one cannot expand the integrand 
in Eq.~(\ref{x_expr}) to arbitrarily high orders
in $\omega_{nA} \, R$ and $\omega_{nB} \, R$,
without incurring infrared divergences for small $x$.
One thus has to introduce a scale-separation 
parameter $\epsilon$, as in Lamb shift 
calculations~\cite{Pa1993,UdKp1996},
to separate the region 
$x \ll \{ \omega_{nA} \, R, \omega_{nB} \, R \}$
from the region
$x \gg \{ \omega_{nA} \, R, \omega_{nB} \, R \}$.
Finally, one obtains the logarithmic coefficient
\begin{eqnarray}
C^{(7,1)}_{1} &=&
- \frac{88}{45} \, \left( 
N_A \, \langle \delta^{(3)}(\vec \rho_B) \rangle +
N_B \, \langle \delta^{(3)}(\vec \rho_A) \rangle 
\right) \,.
\end{eqnarray}
The expression for the accompanying nonlogarithmic term 
is more complicated and of the Bethe logarithm type,
\begin{eqnarray}
\label{C710}
&& \hspace{-2cm} 
C^{(7,0)}_{1} =
\frac{8}{675} \left( 193 - 165 \, \gamma_E \right) \,
\left(
N_A \, \langle \delta^{(3)}(\vec \rho_B) \rangle +
N_B \, \langle \delta^{(3)}(\vec \rho_A) \rangle
\right)
+ \frac{88}{135 \pi} \,
\nonumber\\[0.1133ex]
&& \hspace{-2cm} \times
\left< \rho^i_A \rho^j_B \,
\frac{ (\calH_A - \calE_A)^4 \ln( 2 \alpha | \calH_A - \calE_A | ) -
(\calH_B - \calE_B)^4 \ln( 2 \alpha | \calH_B - \calE_B | )}%
{ (\calH_A - \calE_A)^2 - (\calH_B - \calE_B)^2 } 
P^i_A P^j_B \right> \,.
\end{eqnarray}
For two identical atoms, the denominator 
$(\calH_A - \calE_A)^2 - (\calH_B - \calE_B)^2 $ vanish if,
in a sum-over-states representation,
the same excited intermediate state enters the 
calculation.
However, the numerator in this case also becomes
singular. Numerically, one could treat the 
problem by adding an infinitesimal 
shift to the Hamiltonian of atom $B$, as
in the replacement $\calH_B - \calE_B \to
\calH_B - \calE_B + \eta$,
and considering the limit $\eta \to 0$ at the end 
of the calculation.
Alternatively, 
for two identical atoms with $\calE_A = \calE_B = \calE_0$,
and $| \langle 0 | \vec \rho_A | n_A \rangle|^2 
= | \langle 0 | \vec \rho_B | n_B \rangle|^2 
= | \langle 0 | \vec \rho | n \rangle|^2$ (for $n_A = n_B$),
the contribution of the same-excitation states in both 
atoms yields a contribution
\begin{eqnarray}
\overline C^{(7,0)}_{1} &=& -\frac{44}{135 \pi} 
\sum_n \left\{ \left| \langle 0 | \vec \rho | n \rangle \right|^2 \right\}^2 \,
( \calE_n - \calE_0 )^4 \left[ 1 + 4 \, \ln(2 \alpha | \calE_n - \calE_0 |) 
\right]
\nonumber\\[0.1133ex]
&=& -\frac{44}{135 \pi}
\sum_n \left\{ \left| \langle 0 | \vec P | n \rangle \right|^2 \right\}^2 \,
\left[ 1 + 4 \, \ln(2 \alpha | \calE_n - \calE_0 |)
\right] \,.
\end{eqnarray}
The full $C^{(7,0)}_{1} $ can, in this case, be obtained
by adding the term $\overline C^{(7,0)}_{1}$ to the term 
from Eq.~(\ref{C710}), when the sum over virtual states in the latter
is restricted to virtual states 
with a manifestly different energy for the two atoms.

The above consideration illustrate that in higher orders,
logarithmic terms (both in $\alpha$ as well as in $R$) 
naturally occur in calculations of the Casimir--Polder 
(van der Waals) interaction and need to be taken into account
in a precise analysis of the problem.
Furthermore, we uncover a Bethe-logarithm-like 
structure in the accompanying nonlogarithmic terms.


%
\section*{References}


\begin{thebibliography}{12}%

\bibitem{Casimir-Polder1948}
Casimir H B G  and Polder D 1948 \textit{Phys.~Rev.} 
\textbf{73}\, 360
%
\bibitem{UdCmVdPRL2017} 
Jentschura U D, Adhikari C M  and Debierre V 2017
\textit{Phys.~Rev.~Lett.} \textbf{118}\, 123001
%
\bibitem{MGA1995}
Jamieson M J,  Drake  G W F, and  Dalgarno  A 1995 
\textit{Phys. Rev. A} \textbf{51}\, 3358 
%
\bibitem{MK1996}
Chen M K  and  Chung K T 1996
{\it Phys. Rev. A} \textbf{53} \,1439
%
\bibitem{ZA1997}
Yan Z C,  Dalgarno A, and Babb J  F 1997 
\textit{Phys. Rev. A} \textbf{55}\, 2882
%
\bibitem{Pachucki2005}
Pachucki K  2005 \textit{Phys.~Rev.~A} \textbf{72}\, 062706
%
\bibitem{Power2001}
Power E A 2001 \textit{Eur.~J.~Phys.} \textbf{22}\, 453 
%
\bibitem{ET1995}
Power  E A and  Thirunamachandran T 1995
\textit{Phys. Rev. A} \textbf{51}\, 3660
%
\bibitem{MRA2015}
Donaire M, Guerout R, and Lambrecht A 2015
\textit{Phys. Rev. Lett.} \textbf{115}\, 033201
%
\bibitem{UVCAN2017}
Jentschura U D, Debierre V, Adhikari C M,  Matveev A, and  Kolachevsky N 2017
\textit{Phys. Rev. A 95}\, 022704
%
\bibitem{UjVd2017}
Jentschura U D, Debierre V 2017
\textit{Phys. Rev. A} \textbf{95}\, 042506
%

\bibitem{JeKeAOP2004}
Jentschura U D and Keitel C H 2004
\textit{Ann. Phys. (N.Y.)} \textbf{310}\, 1

%
\bibitem{DPCraig}
Craig D P and Thirunamachandran  T 1984 
\textit{Molecular quantum electrodynamics: An introductin
to radiation-molecule interactions} (Academic Press, New York, NY)
%
\bibitem{Akbarsalam}
Salam  A  2009 \textit{Molecular quantum electrodynamics} 
(John Wiley \& Sons, Inc. Hoboken, NJ, 2009)\,121
%
\bibitem{LandauLifshitz4}
Berestestskii V B, Lifshitz E M  and Pitaevskii  L P 1982
\textit{Quantum Electrodynamics, Volume 4 of the Course
on Theoretical Physics}, 2nd Ed ( Pergamon Press, Oxford, UK)
%
\bibitem{CAThesis2017} 
Adhikari  C M 2017 \textit{Long-range interatomic interactions: 
Oscillatory tails and hyperfine perturbations}, PhD thesis,
Missouri University of Science and Technology, available at 
\url{http://scholarsmine.mst.edu/doctoral_dissertations/2615}
%
%
\bibitem{UdKp1996}
Jentschura U and Pachucki K 1996 \textit{Phys. Rev. A} \textbf{54}\, 1853

\bibitem{OurPRA1from2017}
Adhikari C M, Debierre V, Matveev A,
Kolachevsky N and  Jentschura U D 2017
\textit{Phys. Rev. A} \textbf{95}\, 022703

%
\bibitem{Udj2003}
Jentschura U D 2003 \textit{J.~Phys.~A} \textbf{36}\, L229

\bibitem{CoDa1961}
Cohen M, Dalgarno A 1961 Proc.~Roy.~Soc.~London A \textbf{261} 565.

%
\bibitem{Udj.Con.Acc.Tech1999}
Jentschura U D, Mohr P J, Soff G and Weniger E J 1999
\textit{Comput.~Phys.~Commun.} \textbf{116}\, 28
%
\bibitem{UdjThesis2003}
Jentschura U D 2002 \textit{Quantum electrodynamic bound-state calculations and 
large-order perturbation theory},
Habilitation thesis, Dresden University of Technology, available at 
\url{https://arxiv.org/pdf/hep-ph/0306153}

%
\bibitem{Pa1993}
Pachucki K 1993 \textit{Ann. Phys. (N.Y.)} \textbf{226} 1
\end{thebibliography}
\end{document}